\documentclass[conf, 10pt]{IEEEtran}

\usepackage{cite}
\usepackage{amsmath,amssymb,amsfonts,amsthm, empheq}
\usepackage{algorithm}
\usepackage{algorithmic}
\usepackage{setspace}
\usepackage{graphicx}
\usepackage{textcomp}
\usepackage{xcolor}
\usepackage{mathtools}
\usepackage{graphicx}
\usepackage{booktabs}
\usepackage{makecell}
\usepackage[font={small}]{caption}
\usepackage{subcaption}
\usepackage{cite}
\usepackage{breqn}
\usepackage{mathrsfs}
\usepackage{accents}
\usepackage{acronym}
\usepackage{multirow}
\usepackage{soul}
\usepackage{bm}
\usepackage{url}
\usepackage{wrapfig}
\usepackage{todonotes}
\usepackage{verbatim}
\usepackage{pdfpages}
\usepackage{siunitx}
\usepackage{adjustbox}

\usepackage{pgfplots}
\pgfplotsset{compat=1.18}
\usepgfplotslibrary{statistics}
\usepackage{tikz}
\usepackage{pgfplotstable}
\usepackage{xcolor}

\usetikzlibrary{shapes, shapes.misc, arrows, positioning, calc, backgrounds, angles, quotes, patterns, pgfplots.groupplots, arrows.meta, decorations.markings}

\renewcommand{\thetable}{\arabic{table}}

\DeclareMathOperator*{\argmin}{arg\,min}

\graphicspath{{./figures/}}
\acrodef{prop}[\textit{MIMORPH}]{MIMO Radio Platform for Heterogeneous wireless systems}

\acrodef{abft}[A-BFT]{Association Beamforming Training}
\acrodef{ack}[ACK]{Acknowledge}
\acrodef{adc}[ADC]{Analog-to-Digital Converter}
\acrodef{aoa}[AoA]{Angle of Arrival}
\acrodef{aod}[AoD]{Angle of Departure}
\acrodef{ap}[AP]{Access Point}
\acrodef{amc}[AMC]{Advanced Mezzanine Card}
\acrodef{awv}[AWV]{Antenna Wave Vector}
\acrodef{axi}[AXI]{Advanced eXtensible Interface}
\acrodef{adc}[ADC]{Analog-to-Digital Converter}
\acrodef{ber}[BER]{Bit Error Rate}
\acrodef{bft}[BFT]{Beamforming Training}
\acrodef{bp}[BP]{Beam Pattern}
\acrodef{brp}[BRP]{Beam Refinement Phase}
\acrodef{bss}[BSS]{Blind Source Separation}
\acrodef{cacc}[CACC]{Cross-Antenna Cross-Correlation}
\acrodef{casr}[CASR]{Cross-Antenna Signal Ratio}
\acrodef{cs}[CS]{Compressed Sensing}
\acrodef{cdf}[CDF]{Cumulative Distribution Function}
\acrodef{cef}[CEF]{Channel Estimation Field}
\acrodef{cfo}[CFO]{Carrier Frequency Offset}
\acrodef{cir}[CIR]{Channel Impulse Response}
\acrodef{cfr}[CFR]{Channel Frequency Response}
\acrodef{crb}[CRB]{Cramér-Rao Bound}
\acrodef{crlb}[CRLB]{Cramér-Rao Lower Bound}
\acrodef{csi}[CSI]{Channel State Information}
\acrodef{csirs}[CSI-RS]{CSI-Reference Signal}
\acrodef{cs}[CS]{Compressed Sensing}
\acrodef{cv}[CV]{Constant Velocity}
\acrodef{cnn}[CNN]{Convolutional Neural Network}
\acrodef{cots}[COTS]{Commercial-Off-The-Shelf}
\acrodef{dac}[DAC]{Digital-to-Analog Converter}
\acrodef{dft}[DFT]{Discrete Fourier Transform}
\acrodef{dl}[DL]{Deep Learning}
\acrodef{dma}[DMA]{Direct Memory Access}
\acrodef{dmg}[DMG]{Directional Multi Gigabit}
\acrodef{dmrs}[DMRS]{Demodulation Reference Signal}
\acrodef{srs}[SRS]{Sounding Reference Signal}
\acrodef{dti}[DTI]{Data Transfer Interval}
\acrodef{edmg}[EDMG]{Enhanced Directional Multi Gigabit}
\acrodef{ekf}[EKF]{Extended Kalman Filter}
\acrodef{elu}[ELU]{Exponential-Linear Unit}
\acrodef{fft}[FFT]{Fast Fourier Transform}
\acrodef{fmcw}[FMCW]{Frequency-Modulated Continuous-Wave}
\acrodef{fim}[FIM]{Fisher Information Matrix}
\acrodef{fov}[FOV]{Field-of-View}
\acrodef{ft}[FT]{Fourier Transform}
\acrodef{fr2}[FR2]{Frequency Range 2}
\acrodef{fpga}[FPGA]{Field Programmable Gate Array}
\acrodef{gpio}[GPIO]{General Purpose Input/Output}
\acrodef{gsps}[GSPS]{Giga-Samples per Second}
\acrodef{har}[HAR]{Human Activity Recognition}
\acrodef{ht}[HT]{High Throughput}
\acrodef{ica}[ICA]{Independent Component Analysis}
\acrodef{idft}[IDFT]{Inverse Discrete Fourier Transform}
\acrodef{if}[IF]{Intermediate Frequency}
\acrodef{ifft}[IFFT]{Inverse Fast Fourier Transform}
\acrodef{ifs}[IFS]{Inter-Frame Spacing}
\acrodef{iht}[IHT]{Iterative Hard Thresholding}
\acrodef{ista}[ISTA]{Iterative Shrinkage-Thresholding Algorithm}
\acrodef{isac}[ISAC]{Integrated Sensing And Communication}
\acrodef{iot}[IoT]{Internet of Things}
\acrodef{imu}[IMU]{Inertial Measurement Unit}
\acrodef{jcs}[JCS]{Joint Communication and Sensing}
\acrodef{jm}[JM]{JUMP-MUSIC}
\acrodef{jpdaf}[JPDAF]{Joint Probabilistic Data Association Filter}
\acrodef{lo}[LO]{Local Oscillator}
\acrodef{los}[LoS]{Line-of-Sight}
\acrodef{lbm}[LBM]{Loop-Back Memory}
\acrodef{lora}[LoRa]{Long-Range wide area}
\acrodef{lambda}[LAMBDA]{Least-Squares AMBiguity Decorrelation Adjustment}
\acrodef{ls}[LS]{Least-Squares}
\acrodef{mae}[MAE]{Mean Absolute Error}
\acrodef{mcs}[MCS]{Modulation and Coding Scheme}
\acrodef{md}[$\mu$D]{micro-Doppler}
\acrodef{mimo}[MIMO]{Multiple Input Multiple Output}
\acrodef{mip}[MIP]{Mixed-Integer Programming}
\acrodef{miqp}[MIQP]{Mixed-Integer Quadratic Program}
\acrodef{mmwave}[mmWave]{Millimeter-Wave}
\acrodef{msps}[MSPS]{Mega-Samples per Second}
\acrodef{mu}[MU]{Multiple User}
\acrodef{music}[MUSIC]{MUltiple SIgnal Classification}
\acrodef{nac}[NAC]{Normalized Auto Correlation}
\acrodef{nco}[NCO]{Numerical Controlled Oscillator}
\acrodef{nlos}[NLoS]{Non-Line-of-Sight}
\acrodef{nn}[NN]{Neural Network}
\acrodef{nls}[NLS]{Nonlinear Least-Squares}
\acrodef{ofdm}[OFDM]{Orthogonal Frequency Division Multiplexing}
\acrodef{omp}[OMP]{Orthogonal Matching Pursuit}
\acrodef{per}[PER]{Packet Error Rate}
\acrodef{pdf}[p.d.f.]{Probability Density Function}
\acrodef{phy}[PHY]{Physical Layer}
\acrodef{pl}[PL]{Programmable Logic}
\acrodef{plm}[PLM]{Physical Layer Model}
\acrodef{pov}[POV]{Point-of-View}
\acrodef{ps}[PS]{Processing System}
\acrodef{po}[PO]{Phase Offset}
\acrodef{prf}[PRF]{Pulse Repetition Frequency}
\acrodef{qam}[QAM]{Quadrature Amplitude Modulation}
\acrodef{ransac}[RANSAC]{Random Sample Consensus}
\acrodef{rf}[RF]{Radio Frequency}
\acrodef{rfsoc}[RFSoC]{Radio Frequency System on a Chip}
\acrodef{rcs}[RCS]{Radar Cross-Section}
\acrodef{rmse}[RMSE]{Root Mean Square Error}
\acrodef{rss}[RSS]{Received Signal Strength}
\acrodef{rom}[ROM]{Read Only Memories}
\acrodef{rx}[RX]{receiver}
\acrodef{sc}[SC]{Single Carrier}
\acrodef{sdr}[SDR]{Software Defined Radio}
\acrodef{sinr}[SINR]{Signal-to-Interference-plus-Noise Ratio}
\acrodef{siso}[SISO]{Single Input Single Output}
\acrodef{sls}[SLS]{Sector Level Sweep}
\acrodef{snr}[SNR]{Signal-to-Noise Ratio}
\acrodef{soc}[SoC]{System on a Chip}
\acrodef{spb}[SPB]{Signal Processing Blocks}
\acrodef{srrc}[SRRC]{Square-Root-Raised-Cosine}
\acrodef{ssb}[SSB]{Synchronization Signal Block}
\acrodef{ssr}[SSR]{Super Sample Rate}
\acrodef{sta}[STA]{Station}
\acrodef{std}[STD]{Standard Deviation}
\acrodef{stf}[STF]{Short Training Field}
\acrodef{stft}[STFT]{Short Time Fourier Transform}
\acrodef{su}[SU]{Single User}
\acrodef{tf}[TF]{Time-Frequency}
\acrodef{to}[TO]{Timing Offset}
\acrodef{toa}[ToA]{Time of Arrival}
\acrodef{tdoa}[TDoA]{Time Difference of Arrival}
\acrodef{tx}[TX]{transmitter}
\acrodef{ue}[UE]{User Equipment}
\acrodef{ula}[ULA]{Uniform Linear Array}
\acrodef{usrp}[USRP]{Universal Software Radio Peripheral}
\acrodef{vht}[VHT]{Very High Throughput}
\acrodef{wlan}[WLAN]{Wireless Local Area Network}
\acrodef{wnalp}[WNALP]{Weighted Normalized Auto-correlation Linear Predictor}

\newcommand{\eq}[1]{Eq.~\eqref{#1}}

\newcommand{\fig}[1]{Fig.~\ref{#1}}

\newcommand{\secref}[1]{Section~\ref{#1}}

\newcommand{\mytexttilde}{{\raise.17ex\hbox{$\scriptstyle\mathtt{\sim}$}}}

\IEEEaftertitletext{\vspace{-2.\baselineskip}}

\begin{document}
\bstctlcite{IEEEexample:BSTcontrol}
\pagenumbering{gobble}

\title{Zero-Overhead Unambiguous Velocity Estimation in Multiband ISAC Systems Under Random Traffic}

\author{Aurora Peloso$^{\dag}$,~\IEEEmembership{Graduate Student Member,~IEEE}, Michele Rossi$^{\dag *}$~\IEEEmembership{Senior Member,~IEEE}, \\Jacopo Pegoraro$^{\dag}$,~\IEEEmembership{Member,~IEEE}
\thanks{$^{\dag}$These authors are with the University of Padova, Dept. of Information Engineering. $^{*}$This author is with the University of Padova, Dept. of Mathematics. 
The work presented in this paper is supported by the Multi-X project that has received funding from the Smart Networks and Services Joint Undertaking (SNS JU) under the European Union’s Horizon Europe research and innovation programme under Grant Agreement No 101192521. 
Corresponding author email: \texttt{aurora.peloso@phd.unipd.it}. 
}
}

\maketitle

\begin{abstract}
This paper proposes an original method for estimating the velocity of a target by leveraging the multiband capabilities of modern \acf{isac} systems.
Traditional Doppler estimation relies on regular sampling rates, but \ac{isac} systems often face irregular packet arrival times because they reuse opportunistic communication traffic.
This non-deterministic timing increases the risk of Doppler ambiguity and aliasing, degrading velocity estimation accuracy.
To resolve this, we advocate exploiting frequency diversity across multiple carrier frequencies to observe Doppler shifts without imposing restrictions on packet timing or requiring dedicated sensing overhead.
A multiband velocity estimation problem is here formulated as a mixed-integer quadratic program by utilizing phase differences from all possible pairwise packet combinations.
By integrating at least one unambiguous phase measurement, the system can reconstruct the true target velocity even under sporadic traffic conditions.
Simulation results using realistic traffic traces demonstrate that this approach significantly outperforms multiband likelihood-based and single-band algorithms, with accuracy improving as frequency separation between bands and inter-packet time intervals increase.
This framework provides a zero-overhead solution for robust velocity estimation in dynamic \ac{isac} environments.
\end{abstract}

\IEEEpeerreviewmaketitle
\vspace{-5mm}
\section{Introduction}\label{sec:intro}

Doppler estimation is a cornerstone of \acf{isac} systems, enabling applications from human activity recognition and motion detection to vital sign monitoring~\cite{zhang2021enabling}.
Standard Doppler estimation typically requires channel estimates acquired at a {\it regular rate}~\cite{zhang2021overview}, analogous to the \ac{prf} used in radar systems~\cite{li2010robust}. 
These channel samples are usually processed via \ac{dft} or subspace-based methods such as \ac{music}. 
However, the channel estimation rate imposes a physical constraint on the \ac{isac} system by determining a {\it maximum observable target velocity}~\cite{pegoraro2022sparcs}. 
Any target moving faster than this limit causes aliasing, resulting in Doppler ambiguity and in a significant degradation of the velocity estimation accuracy.

To achieve unambiguous Doppler estimation, traditional strategies generally rely on either {\it (i)} increasing the channel estimation rate, which demands extremely dense communication traffic or the transmission of dedicated sensing packets~\cite{pegoraro2022sparcs}, or {\it (ii)} employing multiple independent, regular sampling rates, as done in \mbox{multi-PRF} radar systems~\cite{younas2020novel, li2010robust}. 
The latter approach exploits different Doppler ambiguities observed across multiple sampling rates to reconstruct an {\it unambiguous} Doppler estimate.

Neither approach is well-suited for \ac{isac} systems, where channel estimation is inherently {\it opportunistic} and contingent upon the random packet arrivals dictated by the network traffic~\cite{ropitault2025open}.
Furthermore, given the wide range of possible target velocities spanning humans, drones, and vehicles, an \ac{isac} system based on regular rates would require constant reconfiguration to remain effective.

\begin{figure}[t!]
    \centering
    \includegraphics[width=\linewidth]{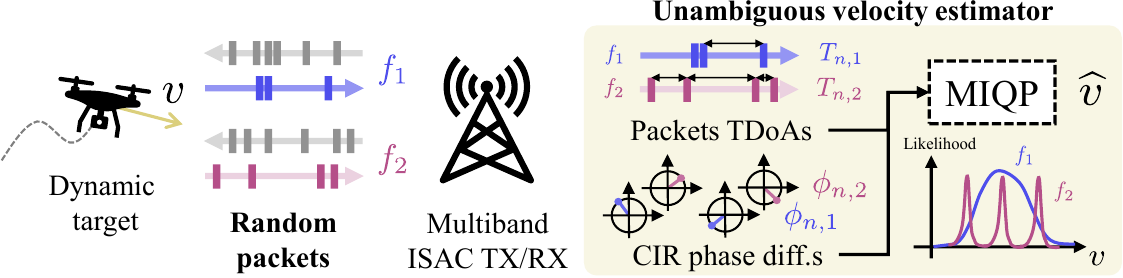}
    \caption{Overview of the proposed method.}
    \label{fig:concept}
\end{figure}

Inspired by the \mbox{multi-PRF} radar concept, the present work proposes an original {\it unambiguous Doppler estimation method} that operates independently of inter-packet arrival times. 
Rather than enforcing impractical temporal regularity, we leverage the \textit{multiband} capabilities of modern communication systems to observe the Doppler shift induced by a target across multiple carrier frequencies, as shown in~\fig{fig:concept}. 
This approach resolves ambiguity without imposing restrictions on the timing of the channel estimation process, by significantly enhancing robustness under realistic and irregular traffic conditions.

Prior literature has addressed traffic irregularity by injecting dedicated sensing packets into the channel~\cite{pegoraro2022sparcs,mason2024using}, but these methods increase system overhead. 
In contrast, our proposed approach leverages Doppler diversity across very distinct frequency bands to eliminate the need for supplemental transmissions, using \ac{tdoa} and phase measurements across packets to formulate velocity estimation as \ac{miqp}.
By decoupling velocity estimation from strict temporal assumptions, we achieve a \textit{zero-overhead} and \textit{unambiguous} Doppler estimation framework that remains reliable even under sporadic traffic patterns.
The frequency bands can span the same range, such as sub-$6$~GHz, or differ significantly, combining, for example, sub-$6$~GHz with millimeter-wave (e.g., $60$~GHz) bands. This flexibility makes the proposed approach broadly applicable across diverse communication technologies.

The idea of exploiting multiple frequency bands in \ac{isac} has been explored to improve the ranging resolution~\cite{li2025enabling} or the localization accuracy~\cite{bedin2025millimeter}.
However, to the best of our knowledge, no prior work in the literature exploits multiband processing for unambiguous velocity estimation.

The contributions of this work are summarized as follows.

(1)~We propose an unambiguous target velocity estimation method that combines channel state information obtained at random time instants across multiple frequency bands.
Our approach entails zero overhead on communication and poses no constraints on the timing of the channel estimation process.

(2)~We formulate and solve the multiband target velocity estimation problem as a \ac{miqp}, utilizing phase differences in the channel estimates obtained in different bands. 

(3)~We numerically evaluate the proposed algorithm, demonstrating that it achieves $10^{-3}$ to $10^{-4}$ relative estimation error in typical conditions, and that it outperforms single-band techniques and a likelihood-based benchmark.

\section{System model}
\label{sec:sys-model}

This section introduces the system model, namely, the \ac{isac} scenario, the channel model, and the phase measurements.
\subsection{Scenario}\label{sec:scenario}
Let us consider a multiband monostatic \ac{isac} transceiver operating across $Q$ distinct frequency bands, with carrier frequencies $f_q$, and bandwidths $B_q$, for \mbox{$q=1, \dots, Q$}.
Each band receives $N_q$ packets, where the receiving time of packet $i=1,\dots,N_q$ is referred to as $T_{i,q}$, which is a random variable (r.v.) depending on the traffic pattern of the \ac{isac} system.
For each received packet, the system uses the preamble part of the packet to obtain an estimate of the \ac{cir}.
Indeed, the proposed method reuses packets originally exchanged for communication purposes rather than relying on dedicated transmissions. 

We consider $K$ scattering points in a $2$-dimensional environment, indexed by $k=1,\dots,K$, and each moving with a velocity vector $\mathbf{v}_k$ containing the velocity components on the $x$ and $y$ axes.
The movement of the $k$-th point relative to the monostatic device introduces a Doppler shift with respect to the original carrier frequency $f_q$. Denoting by $c$ the speed of light, the Doppler frequency of the $k$-th point observed in the $q$-th band is $f_{k, q}^{\rm D} = 2 v_k f_q/c$,
where $v_k$ is the radial velocity of the $k$-th scattering point, i.e., 
$v_k = \|\mathbf{v}_k\| \cos \gamma_k$, where $\gamma_k$ is the angle between the velocity vector and the segment connecting the scattering point to the monostatic receiver.

In the remainder, we analyze the Doppler-induced phase and frequency variations observed across the received signal packets to estimate the target’s \textit{radial} velocity $v_k$.
Estimating the velocity vector $\mathbf{v}_k$ in general requires a multistatic setting with multiple receivers, and it is not addressed here.

\subsection{Channel model}

We consider a multipath channel where the $K$ scattering points are organized into $L$ sensing targets. 
Each target $\ell \in \left\{1, \cdots, L  \right\}$ contributes with $M_\ell$ propagation paths due to its potentially different moving parts, with~\mbox{$K = \sum_{\ell = 1}^{L}M_{\ell}$}. 

The \ac{cir} in frequency band $q$ is modeled as
\begin{align}
\label{eq:cir_multiple_component}
h_q(t,\tau) =& \sum_{\ell=1}^{L} \sum_{m=1}^{M_\ell}\beta_{\ell, m, q} e^{ j 2\pi f_{\ell, m, q}^{\rm D} t } 
\chi_q(\tau - \tau_{\ell, m})
+ w_q(t, \tau) ,
\end{align}
where $\chi_q(x) = \mathrm{sinc}\left(B_qx\right) = \sin (\pi B_qx)/(\pi B_q x)$ for $x\neq 0$ and $1$ elsewhere. 
In~\eq{eq:cir_multiple_component}, $t$ is the so-called \textit{slow time} index, and indicates the reception time instant of the packet from which the \ac{cir} is estimated. 
The \textit{fast time} index is $\tau$ and captures the multipath delay profile. 
$\tau_{\ell, m}$ is the propagation delay of the component due to the $m$-th part of the $\ell$-th target. 
$w_q(t,\tau)$ is a r.v. describing the noise on the \ac{cir}, which is assumed to be a complex Gaussian process with variance $\sigma^2_{w_q}$. We underline that although the model in \eq{eq:cir_multiple_component} is formulated for passive sensing, the proposed velocity estimation method can also be applied to active devices.

The contribution of each target to the \ac{cir} is the result of the superposition of $M_{\ell}$ components $m$ with complex amplitude $\beta_{\ell,m,q}$ and Doppler frequency $f_{\ell, m, q}^{\rm D}$.
We assume that the components of each target cannot be distinguished by the receiver in the delay domain, since the distance between the peaks falls below the delay resolution of the \ac{isac} system, which is $\Delta \tau_q = 1 / B_q$, where $B_q$ represents the bandwidth of the $q$-th frequency band. 
For typical values of the bandwidth of \ac{isac} systems, individual components appear as a single observable peak with amplitude arising from the aggregated contribution of all the individual components (indexed via variable $m$ in \eq{eq:cir_multiple_component}). 
In this study, we assume that the bandwidths of the $Q$ considered sub-bands are comparable, i.e., they provide similar delay resolutions $\Delta \tau_q$ and can thereby detect the same set of targets $\ell$ in the environment.
\vspace{-2mm}

\subsection{Phase measurements model}\label{sec:phase-model}

Despite the presence of $L$ targets in the \ac{cir}, our method can be applied independently to each of them after they have been separated and tracked using existing approaches~\cite{zhang2021overview}.

Therefore, without loss of generality, we focus on a single target $\ell$, and restrict the analysis to its dominant kinematic component (e.g., the motion of the torso in a human mobility scenario).
The assumption of a single target component is lifted in~\secref{sec:results} to demonstrate that our approach is robust to multiple-component targets.

In the single-target $\ell$, single-component scenario, \eq{eq:cir_multiple_component} simplifies to
\begin{equation}
    h_q(t,\tau) 
= \beta_{\ell,q} e^{ j 2\pi f_{\ell,q}^{\rm D} t  }\,\chi_q\big(\tau - \tau_\ell\big) + w_q(t, \tau).
\label{eq:cir_single_component}
\end{equation}
In the following, we assume that processing is performed in a short, coherent time interval in which the amplitude and phase of $\beta_{\ell, q}$ do not change significantly.
Moreover, we drop the index $\ell$ of the target to simplify the notation.

The target velocity $v$ is embedded in the received signal phase, inside the Doppler frequency $f^{\rm D}_{q}$. 
Hence, to estimate $v$, the algorithm extracts the \ac{cir} from each band $q$ and received packet at time $T_{i,q}$ and computes its instantaneous phase
\begin{equation}
\label{eq:phase_arctan}
 \psi_{i,q} = \mathrm{atan2} \left(   \Im\left\{h_q(T_{i,q}, \tau)\right\},    \Re \left\{h_q(T_{i,q}, \tau) \right\}\right),
\end{equation}
where $\mathrm{atan2}$ is the 2-argument arctangent function and $\Re$, $\Im$ denote the real and imaginary parts of a complex number, respectively. 
Given the reception of two packets at \acp{toa} $T_{i,q}$ and $T_{j,q}$, with $T_{i,q}<T_{j,q}$ the corresponding instantaneous phase difference $n$ is 
\begin{equation} \label{eq:phase_diff_equation}
 \phi_{n,q} = \psi_{i,q} - \psi_{j,q}, 
\end{equation}
and lies in the interval $[0,2\pi]$. 
However, the actual \textit{physical} phase evolution over the \ac{tdoa} \mbox{$\Delta T_{n,q} = T_{i,q}-T_{j,q}$}, is
\begin{equation}
     \theta_{n,q} = \frac{4 \pi v f_q \Delta T_{n,q}}{c},
\end{equation} 
which is an unbounded quantity.

Note that the receiver only observes the \textit{wrapped} phase 
\begin{equation}
    \label{eq:phase_ambiguity}
     \phi_{n,q} = (\theta_{n,q} + w_{\phi_{n,q}})\mod{2\pi},
\end{equation} 
with $w_{\phi_{n,q}}$ representing the noise in the phase measurement. \eq{eq:phase_ambiguity} leads to the \textit{phase ambiguity problem} where the total true phase increment can only be recovered up to an unknown integer multiple of $2 \pi$. 

Assuming a sufficiently high \ac{snr} for the estimated \ac{cir}, in equation \eq{eq:phase_arctan}, the $\mathrm{atan2}$ function can be linearized. 
Under this approximation, the noise r.v. $w_{\phi_{n,q}}$ can be modeled as Gaussian. 
This assumption is not restrictive, as it does not imply a high communication \ac{snr}, but rather a high estimation \ac{snr} for the \ac{cir}. 
Channel estimation techniques achieve this by coherently combining multiple pilot symbols, thereby reducing the noise variance.

Due to the phase ambiguity, we model $\theta_{n,q}$ as
\begin{equation}
\label{eq:total_phase_evolution}
         \theta_{n,q}=  \phi_{n,q} + w_{\phi_{n,q}}+ 2 \pi R_{n,q} , \quad  R_{n,q} \in \mathbb{Z}, \\ 
\end{equation}
where $R_{n,q}$ is an integer representing the number of complete phase rotations (i.e., complete revolutions in the I/Q plane) that occur between the two packet receptions.
\vspace{-3mm}


\subsection{Maximum unambiguous velocity and resolution}

Conventional approaches for Doppler frequency estimation process the \ac{cir} with the \ac{dft} or superresolution methods such as \ac{music}~\cite{zhang2021overview}. 
This enables identifying the Doppler frequency component, which can be used to estimate the velocity of the target. 
A key requirement of \ac{dft}-based methods is that the received signal needs to be sampled on a uniform temporal grid, to ensure consistency with discrete-time Fourier analysis.
The key parameters governing a velocity measurement system are the maximum velocity observable $v_{\max, q}$ and the velocity resolution $\Delta v_q$ associated with the $q$-th band, expressed as~\cite{pegoraro2022sparcs}
\begin{equation}
    v_{\max, q} = \frac{c}{4 f_q  \Delta T_q}, \quad
    \Delta v_q = \frac{c}{2 f_q  N_q \Delta T_q}, 
\end{equation}
where $\Delta T_q$ is the time between the reception of two consecutive packets (assuming constant inter-packet time) and $N_q$ is the number of available packets in band $q$.

The proposed method overcomes the limitations of single-band systems deriving from the maximum measurable velocity $v_{{\max}, q}$ and its resolution $\Delta v_q$. 
For illustration purposes, think of a system operating in two different bands: one with a low carrier frequency $f_{1}$ and the other with a high frequency $f_{2}$. 
The rationale is that at $f_{1}$ the system attains a large range of detectable velocities (as $v_{\max,1}$ will be large), while at $f_{2}$ it attains a high velocity resolution (as $\Delta v_2$ will be small). 
These qualities can be jointly exploited to enhance the velocity estimation accuracy. 

\section{Methodology}\label{sec:method}

Given the $N_q$ received packets for frequency $q$, the total number of computable phase differences $\phi_{n,q}$ is $P_q$, which amounts to all the possible pairwise packet combinations $ P_q = N_q (N_q-1)/2$.
Placing the \acf{toa}s $T_{n,q}$ in ascending order, for a given band $q$, the phase differences $\phi_{n,q}$ are obtained as specified in \eq{eq:phase_diff_equation}
for all possible $(i,j)$ packet pairs, i.e., $\quad 1 \le i < j \le N_q$ and $n \in \left\{1, \dots, P_q\right\}$. This approach maximizes the number of available measurements, thereby increasing the measurement redundancy and improving the reliability of the final velocity estimation.
Each phase difference $\phi_{n,q}$ is associated with a \ac{tdoa} $\Delta T_{n,q}$. It is worth noting that the proposed method does not impose any constraint on the structure of $\Delta T_{n,q}$: times $T_{i,q}$ and $T_{j,q}$ are neither required to be equally spaced, nor to be the same across different bands. This explains the dependency of $\Delta T_{n,q}$ on the phase difference measurement $n$ and on the specific band $q$.

\noindent \eq{eq:total_phase_evolution} is thus rewritten for each phase difference $\phi_{n,q}$ as
 \begin{equation}
    \phi_{n,q} = 4 \pi \frac{v}{c} f_{q} \Delta T_{n,q} - 2 \pi R_{n,q} + w_{\phi_{n,q}}.
  \label{eq:general_phase_difference}
 \end{equation}
Considering $P_q$ phase-difference measurements at each frequency $f_q$, the number of the unknown integers $R_{n,q}$ is $N = \sum_{q=1}^Q P_q$. When the unknown velocity is also included, the total number of unknowns becomes $N+1$, while the number of equations remains $N$. The resulting system is therefore underdetermined. 

To resolve this, we require that {\it at least one unambiguous phase measurement} is present among the $N$ available. 
Without loss of generality, we identify the unambiguous phase measurement as the \textit{first} available one at the \textit{lowest} carrier frequency,  which we assume to have index $q=1$ without loss of generality, i.e., $f_{1} = \min (f_{1},\cdots,f_{Q})$.
The unambiguous phase measurement is therefore $\phi_{1, 1}$.
The corresponding total accumulated phase, $\theta_{1,1}$, in the time interval $\Delta T_{1,1}$ remains strictly smaller than~$\pi$. 
Note that requiring that the phase remains smaller than $2 \pi$ is not enough, as this phase would still be ambiguous in terms of the direction of rotation. 

The reason why we consider the lowest carrier frequency is explained as follows.
Taking the noise term into account, our requirement amounts to $\theta_{1, 1} + w_{\phi_{1,1}} < \pi$. 
The noise term $w_{\phi_{1,1}}$ is here upper bounded by $3\sigma_{w_{\phi}}$, as for a Gaussian r.v. this range encompasses approximately $99.73$\% of the possible values. 
The inequality is thus rewritten as
\begin{equation} \label{eq:T_max_1}
\theta_{1,1}+ 3 \sigma_{w_{\phi}} = 4 \pi \frac{v}{c} f_{1} \Delta T_{1,1} + 3 \sigma_{w_{\phi}} <\pi.
\end{equation}
Solving the last inequality for $\Delta T_{1,1}$ leads to 
\begin{equation}
\label{eq:T_max_1}
    \Delta T_{1,1} < \frac{c(\pi -3 \sigma_{w_\phi})}{4 \pi f_1 v}.
\end{equation} 
From \eq{eq:T_max_1}, one can see that the lowest carrier frequency across the available subsystems translates into the largest $\Delta T_{1,1}$ that is admitted for our measurement. 
This means that the lowest carrier frequency is more likely to observe an unambiguous measurement.
For the velocity $v$ in \eq{eq:T_max_1}, we use the maximum velocity that we need to track from the observed physical environment, which we refer to as $v_{\max}$.
Keeping these facts into account, \eq{eq:T_max_1} becomes
\begin{equation}
\label{eq:T_max_2}
    \Delta T_{1,1} < \frac{c(\pi -3 \sigma_{w_\phi})}{4 \pi f_{1} v_{\max}}.
\end{equation}
We remark that the assumption of having one unambiguous measurement is reasonable, given practical values of the packet \ac{tdoa} and target velocities. As an example, in the real traffic data we use in our evaluation in~\secref{sec:results}~\cite{phillips2007pdx}, using $\Delta T_{1,1} \leq 0.5$~ms, which occurs frequently, and $f_1=2.4$~GHz, the system would be able to unambiguously observe velocities up to $v_{\max} \approx 220$~km/h. 

To solve the considered velocity estimation problem, we reformulate equation \eq{eq:general_phase_difference} in matrix form. The $N$-dimensional vector of measurements is composed of all the available phase differences $\phi_{n,b}$, with $n \in 1, \dots, P_q$ and for all bands $q$ is 
\begin{equation}
\label{eq:y_equation}
\mathbf{y} = \left[ \phi_{1,1},\!\ldots\!,\phi_{P_1,1}, \phi_{1,2},\!\ldots\!,\phi_{P_2,2},\!\ldots\!, \phi_{1,Q},\!\ldots\!,\phi_{P_Q,Q} \right]^T
\end{equation}
The first element of $\mathbf{y}$ is the unambiguous phase measure acquired at $f_1$ in the time interval $\Delta T_{1,1}$.
We further define a diagonal matrix $\mathbf{A}$ containing the velocity coefficients  in the first term of \eq{eq:general_phase_difference}, 
\begin{equation}\label{eq:A_equation}
\begin{split}
\mathbf{A}
=
& \frac{4\pi}{c}\, \mathrm{diag}
(\Delta T_{1,1} f_1 ,
\dots
  \Delta T_{P_1,1} f_1,
  \Delta T_{1,2} f_2 ,
 \\
&   \dots,
 \Delta T_{P_2,2} f_2, \dots,\Delta T_{1,Q}f_Q,\, \dots , \Delta T_{P_Q,Q}f_Q).
\end{split}
\end{equation}
where $\Delta T_{n,q}$ correspond to the \ac{tdoa} for which the phase difference $\phi_{n,q}$ is computed.
A vector $\mathbf{r}$ contains the unknown integers $R_{n,q}$ of \eq{eq:general_phase_difference} 
\begin{equation}
\mathbf{r} = \left[ 0,\!\ldots\!,R_{P_1,1}, R_{1,2},\!\ldots\!,R_{P_2,2},\!\ldots\!, R_{1,Q},\!\ldots\!,R_{P_Q,Q} \right]^T.
\end{equation}
Note that the first element of the vector $\mathbf{r}$ is zero since it corresponds to the unambiguous measurement acquired at $\Delta T_{1,1}$, hence reducing the number of unknowns to $N$. 
The phase noise vector is defined as:
\begin{equation}
\mathbf{w} = \left[w_{1,1},\!\ldots\!,w_{P_1,1},w_{1,2},\!\ldots\!,w_{P_2,2},\!\ldots\!,w_{1,Q},\!\ldots\!,w_{P_Q,Q}\right]^T.
\end{equation}
The final $N\times N$ system of equations is
\begin{equation}
\label{eq:system_matrix_form}
    \mathbf{y} =  v\mathbf{A}\mathbf{1}  - 2 \pi\mathbf{r} + \mathbf{w},
\end{equation}
where $\mathbf{1}$ is a vector of ones of dimension~$N$.
Note that using all the possible phase differences $P_q$ for each band $q$, we increase the number of equations in \eq{eq:system_matrix_form}, improving the estimation accuracy. 
The optimal velocity $\widehat{v}$ and vector $\widehat{\mathbf{r}}$ are thus found solving the following \textit{integer least-squares} problem,
\begin{equation}
\label{eq:least-square}
    \argmin_{v, \mathbf{r}} \; \left\| \mathbf{y} - v\mathbf{A}\mathbf{1} + 2 \pi \mathbf{r} \right\|_2^2 , 
    \quad v \in \mathbb{R}, \; \mathbf{r} \in \mathbb{Z}^{N}.
\end{equation}
The optimization problem in~\eq{eq:least-square} is a \ac{miqp} that can be solved with existing methods, e.g.,
via a branch-and-bound algorithm extended to handle quadratic objective functions. 
Although \textit{integer least-squares} problems are NP-hard, algorithms such as LAMBDA~\cite{teunissen1997performance} provide computationally efficient solutions, supporting real-time applications. 


\section{Numerical results}\label{sec:results}

This section presents the simulation results for the proposed algorithm,
starting from the setup, then analyzing single- and multi-component targets
scenarios, and finally comparing it against two benchmarks.

\textbf{Simulation setup.} The input measurements consist of complex values derived from the peak of the estimated \ac{cir} corresponding to the target of interest. 
The phase measurement noise, $w_{\phi}$, is modeled as a zero-mean Gaussian random variable, \mbox{$w_{\phi} \sim \mathcal{N}(0, \sigma_{w_{\phi}}^{2})$}. 
The simulation considers a two-band system with $f_{1}=2.4$~GHz and $f_2 \in \{5, 7, 14, 28, 60\}$~GHz. 

The amplitude coefficient from \eq{eq:cir_single_component} is $\beta_q = \tilde\beta C_q$, where
$C_{q} \sim \mathcal{N}(1, \sigma_{C}^{2})$ with $\sigma_{C}=0.05$ models frequency-dependent variations, and $\tilde\beta \sim \mathcal{U}(0,1]$ is the scattering coefficient.
The target velocity is $v \sim \mathcal{U}[-50, 50]$~m/s, chosen to jointly evaluate estimation accuracy and robustness to ambiguity.

\ac{toa} values are drawn from empirical Wi-Fi traffic traces provided in~\cite{phillips2007pdx},
with packet \acp{tdoa} from $0.0385$~ms to $0.15$~s.
Given the large number of packets in the traces, for our analysis \ac{toa} values were selected uniformly among the recorded ones in the dataset within the interval $[T_{\min}, T_{\max}]$, where $T_{\min} = 0.0385$ ms and $T_{\max} \in \{0.1, 1, 10, 57\}$ ms. 
Unless stated otherwise, $N_1 = N_2 = 4$ packets per band are used, giving $P_{q} = 12$ pairwise phase differences per band via \eq{eq:phase_diff_equation}. 
Note that this choice is not restrictive, as a different number of packets could be used in each band.
The velocity is estimated by solving \eq{eq:least-square}, and accuracy is measured by the relative error $\varepsilon_{v} = |\hat{v} - v| / |v|$.

\begin{figure}[t!]
    \centering
    \subfloat{
        \resizebox{0.4\textwidth}{!}{\begin{tikzpicture}
\definecolor{color1}{RGB}{102,104,238}
\definecolor{color2}{RGB}{204,187,68}
\definecolor{color3}{RGB}{170,51,119}
\definecolor{color4}{RGB}{187,187,187}
\begin{axis}[
    width=0cm,
    height=0cm,
    axis line style={draw=none},
    tick style={draw=none},
    scale only axis,
    xmin=0, xmax=1,
    ymin=0, ymax=1,
    xtick=\empty,
    ytick=\empty,
    axis background/.style={fill=white},
    legend style={
        legend cell align=left,
        fill opacity=1,
        align=center,
        draw=white,
        font=\small,
        at={(0.5,0.5)},
        anchor=center,
        /tikz/every even column/.append style={column sep=0.8em} 
    },
    legend columns=4, 
]
\addlegendimage{area legend, fill=color1, draw=black, fill opacity = 0.7}
\addlegendentry{$T_{\max} = 0.1$~ms}

\addlegendimage{area legend, fill=color2, draw=black, fill opacity = 0.7}
\addlegendentry{$T_{\max}= 1$~ms}

\addlegendimage{area legend, fill=color3, draw=black, fill opacity = 0.7}
\addlegendentry{$T_{\max} = 10$~ms}

\addlegendimage{area legend, fill=color4, draw=black, fill opacity = 0.7}
\addlegendentry{$T_{\max} = 57$~ms}

\end{axis}
\end{tikzpicture}}
    } \\
    \setcounter{subfigure}{0}
    \subfloat[$\sigma_{w_\phi} = 10^\circ$.  \label{fig:results_multiband_alg_a}]{
        \resizebox{0.22\textwidth}{!}{
\begin{tikzpicture}
\definecolor{color1}{RGB}{102,104,238}
\definecolor{color2}{RGB}{204,187,68}
\definecolor{color3}{RGB}{170,51,119}
\definecolor{color4}{RGB}{187,187,187}
\begin{axis}[
  boxplot/draw direction=y,
  ylabel={Error $\varepsilon_v$},
  xlabel={Band 2 [GHz]},
  ylabel shift=-5 pt,
  ymin=1e-7, ymax=1,
  xmin=-0.5, xmax=4.5,
  ymode = log,
  yticklabel style={/pgf/number format/fixed},
  ytick={0.1, 0.01, 0.001, 0.0001, 0.00001, 0.000001, 0.0000001},
  ymajorgrids,
  xtick={0,1,2,3,4},
  xticklabels={$5.0$,$7.0$,$14.0$,$28.0$,$60.0$},
  xlabel style={font=\small}, ylabel style={font=\small}, ticklabel style={font=\small},
  /pgfplots/boxplot/box extend=0.2,
  height=5cm,
  width=7cm,
  boxplot/every box/.style={solid, draw=black},
  boxplot/every whisker/.style={solid, black},
  boxplot/every cap/.style={solid, black},
  boxplot/every median/.style={solid, black},
]
  \addplot+[fill=color1, fill opacity=0.7, draw=black, boxplot prepared={
    median=0.09779707451053268,
    upper quartile=0.2998097043920485,
    lower quartile=0.0336955019999998,
    upper whisker=0.6968354551854563,
    lower whisker=4.955708477714866e-05,
    draw position=-0.30000000000000004
  }] coordinates {};

  \addplot+[fill=color1, fill opacity=0.7, draw=black, boxplot prepared={
    median=0.1293216899717855,
    upper quartile=0.4033435977150638,
    lower quartile=0.04537577904213816,
    upper whisker=0.927751315365127,
    lower whisker=7.855903891189723e-05,
    draw position=0.7
  }] coordinates {};

  \addplot+[fill=color1, fill opacity=0.7, draw=black, boxplot prepared={
    median=0.1333116947138248,
    upper quartile=0.34410022106855553,
    lower quartile=0.05134529422885671,
    upper whisker=0.7823196129340997,
    lower whisker=0.00011299893496856594,
    draw position=1.7000000000000002
  }] coordinates {};

  \addplot+[fill=color1, fill opacity=0.7, draw=black, boxplot prepared={
    median=0.12588098685112858,
    upper quartile=0.2758003053483186,
    lower quartile=0.0513691905207795,
    upper whisker=0.6103177132505925,
    lower whisker=0.00031930692787112937,
    draw position=2.7
  }] coordinates {};

  \addplot+[fill=color1, fill opacity=0.7, draw=black, boxplot prepared={
    median=0.07925581542474797,
    upper quartile=0.17204399446535001,
    lower quartile=0.03472812180752916,
    upper whisker=0.3676955488005527,
    lower whisker=0.0001682481713171964,
    draw position=3.7
  }] coordinates {};

  \addplot+[fill=color2, fill opacity=0.7, draw=black, boxplot prepared={
    median=0.06710862082568442,
    upper quartile=0.1683425942997776,
    lower quartile=0.029118382062936,
    upper whisker=0.3745736144865549,
    lower whisker=0.0002389693497375069,
    draw position=-0.09999999999999998
  }] coordinates {};

  \addplot+[fill=color2, fill opacity=0.7, draw=black, boxplot prepared={
    median=0.04863714702143872,
    upper quartile=0.10258418489939247,
    lower quartile=0.021985009482273485,
    upper whisker=0.22290098916729073,
    lower whisker=8.703095097373875e-05,
    draw position=0.9
  }] coordinates {};

  \addplot+[fill=color2, fill opacity=0.7, draw=black, boxplot prepared={
    median=0.027269853732218408,
    upper quartile=0.0611039617615872,
    lower quartile=0.011106333817662432,
    upper whisker=0.13507040875951504,
    lower whisker=8.186224973136003e-05,
    draw position=1.9000000000000001
  }] coordinates {};

  \addplot+[fill=color2, fill opacity=0.7, draw=black, boxplot prepared={
    median=0.014020532784122436,
    upper quartile=0.0348785624005152,
    lower quartile=0.0058940092779320484,
    upper whisker=0.07756885292459569,
    lower whisker=3.976770437298476e-06,
    draw position=2.9000000000000004
  }] coordinates {};

  \addplot+[fill=color2, fill opacity=0.7, draw=black, boxplot prepared={
    median=0.007395704628980858,
    upper quartile=0.018085011407214708,
    lower quartile=0.0030086389052010457,
    upper whisker=0.040480537200129216,
    lower whisker=2.9756517364002053e-06,
    draw position=3.9000000000000004
  }] coordinates {};

  \addplot+[fill=color3, fill opacity=0.7, draw=black, boxplot prepared={
    median=0.008267680728168324,
    upper quartile=0.020918872501692984,
    lower quartile=0.003322570886727149,
    upper whisker=0.04672011175877483,
    lower whisker=3.489091990461566e-07,
    draw position=0.09999999999999998
  }] coordinates {};

  \addplot+[fill=color3, fill opacity=0.7, draw=black, boxplot prepared={
    median=0.006247658506753228,
    upper quartile=0.015669935676273117,
    lower quartile=0.002365231165688198,
    upper whisker=0.03543240892335297,
    lower whisker=2.471938014210381e-07,
    draw position=1.1
  }] coordinates {};

  \addplot+[fill=color3, fill opacity=0.7, draw=black, boxplot prepared={
    median=0.00347431734031045,
    upper quartile=0.00865523195038286,
    lower quartile=0.0014610647595435411,
    upper whisker=0.019372215670893637,
    lower whisker=1.5587496467074467e-06,
    draw position=2.1
  }] coordinates {};

  \addplot+[fill=color3, fill opacity=0.7, draw=black, boxplot prepared={
    median=0.0018853408196927296,
    upper quartile=0.004903707241199128,
    lower quartile=0.0007571549200642618,
    upper whisker=0.010891898059775036,
    lower whisker=3.2125558297185886e-06,
    draw position=3.1
  }] coordinates {};

  \addplot+[fill=color3, fill opacity=0.7, draw=black, boxplot prepared={
    median=0.001015783744476956,
    upper quartile=0.0029955048642536413,
    lower quartile=0.00036377826081198284,
    upper whisker=0.006865303616531478,
    lower whisker=7.530291473011767e-07,
    draw position=4.1
  }] coordinates {};

  \addplot+[fill=color4, fill opacity=0.7, draw=black, boxplot prepared={
    median=0.0007942694126759071,
    upper quartile=0.002310815017750662,
    lower quartile=0.00036814208868943184,
    upper whisker=0.00522074573498139,
    lower whisker=5.1403664313985e-07,
    draw position=0.30000000000000004
  }] coordinates {};

  \addplot+[fill=color4, fill opacity=0.7, draw=black, boxplot prepared={
    median=0.0007225201706235215,
    upper quartile=0.002092558818822003,
    lower quartile=0.00028662163311118883,
    upper whisker=0.004720492067791785,
    lower whisker=2.3997484002769843e-06,
    draw position=1.3
  }] coordinates {};

  \addplot+[fill=color4, fill opacity=0.7, draw=black, boxplot prepared={
    median=0.00043611093067130307,
    upper quartile=0.0014316054065102069,
    lower quartile=0.00017637425006383306,
    upper whisker=0.003155745658605426,
    lower whisker=9.5521571244276e-07,
    draw position=2.3000000000000003
  }] coordinates {};

  \addplot+[fill=color4, fill opacity=0.7, draw=black, boxplot prepared={
    median=0.00024406477969177447,
    upper quartile=0.0012414059614899305,
    lower quartile=8.396625581219836e-05,
    upper whisker=0.002942490776762759,
    lower whisker=2.0085657029566602e-07,
    draw position=3.3000000000000003
  }] coordinates {};

  \addplot+[fill=color4, fill opacity=0.7, draw=black, boxplot prepared={
    median=0.00014401387271242364,
    upper quartile=0.0013826242918139036,
    lower quartile=4.611971216590009e-05,
    upper whisker=0.0033756822797758705,
    lower whisker=2.018114847447252e-07,
    draw position=4.3
  }] coordinates {};
\end{axis}
\end{tikzpicture}}
    }
    \subfloat[$\sigma_{w_\phi} = 20^\circ$. \label{fig:results_multiband_alg_b}]{
        \resizebox{0.22\textwidth}{!}{
\begin{tikzpicture}
\definecolor{color1}{RGB}{102,104,238}
\definecolor{color2}{RGB}{204,187,68}
\definecolor{color3}{RGB}{170,51,119}
\definecolor{color4}{RGB}{187,187,187}
\begin{axis}[
  boxplot/draw direction=y,
  ylabel={Error $\varepsilon_v$},
  xlabel={Band 2 [GHz]},
  ylabel shift=-5 pt,
  ymin=1e-7, ymax=1,
  ymode = log,
  xmin=-0.5, xmax=4.5,
  yticklabel style={/pgf/number format/fixed},
  ymajorgrids,
    ytick={0.1, 0.01, 0.001, 0.0001, 0.00001, 0.000001, 0.0000001},
  xtick={0,1,2,3,4},
  xticklabels={$5.0$,$7.0$,$14.0$,$28.0$,$60.0$},
  xlabel style={font=\small}, ylabel style={font=\small}, ticklabel style={font=\small},
  /pgfplots/boxplot/box extend=0.2,
  height=5cm,
  width=7cm,
  boxplot/every box/.style={solid, draw=black},
  boxplot/every whisker/.style={solid, black},
  boxplot/every cap/.style={solid, black},
  boxplot/every median/.style={solid, black},
]
  \addplot+[fill=color1, fill opacity=0.7, draw=black, boxplot prepared={
    median=0.20762499798424355,
    upper quartile=0.6245425689411208,
    lower quartile=0.07076381338850896,
    upper whisker=1.4156147648648376,
    lower whisker=0.00012430824606574656,
    draw position=-0.30000000000000004
  }] coordinates {};

  \addplot+[fill=color1, fill opacity=0.7, draw=black, boxplot prepared={
    median=0.22238476432361565,
    upper quartile=0.6459390140995058,
    lower quartile=0.07471103509129319,
    upper whisker=1.496443170044546,
    lower whisker=0.00014348782250078377,
    draw position=0.7
  }] coordinates {};

  \addplot+[fill=color1, fill opacity=0.7, draw=black, boxplot prepared={
    median=0.2764066820814861,
    upper quartile=0.681323327351953,
    lower quartile=0.10930138964149017,
    upper whisker=1.5355263421855139,
    lower whisker=0.0014572239578355044,
    draw position=1.7000000000000002
  }] coordinates {};

  \addplot+[fill=color1, fill opacity=0.7, draw=black, boxplot prepared={
    median=0.2685731592319056,
    upper quartile=0.631884365966029,
    lower quartile=0.11755959508696906,
    upper whisker=1.4019358913445403,
    lower whisker=5.737614425135263e-05,
    draw position=2.7
  }] coordinates {};

  \addplot+[fill=color1, fill opacity=0.7, draw=black, boxplot prepared={
    median=0.15720965624251226,
    upper quartile=0.33968497629494976,
    lower quartile=0.06897308751620944,
    upper whisker=0.7428211702984714,
    lower whisker=0.000283409533807922,
    draw position=3.7
  }] coordinates {};

  \addplot+[fill=color2, fill opacity=0.7, draw=black, boxplot prepared={
    median=0.13732147630015393,
    upper quartile=0.315484442753345,
    lower quartile=0.058067114737043915,
    upper whisker=0.6998619320288317,
    lower whisker=0.0010360272852564805,
    draw position=-0.09999999999999998
  }] coordinates {};

  \addplot+[fill=color2, fill opacity=0.7, draw=black, boxplot prepared={
    median=0.10556331943692168,
    upper quartile=0.23078136075384786,
    lower quartile=0.0469856323036598,
    upper whisker=0.49849460185983485,
    lower whisker=0.00014260227222077929,
    draw position=0.9
  }] coordinates {};

  \addplot+[fill=color2, fill opacity=0.7, draw=black, boxplot prepared={
    median=0.06099305198740143,
    upper quartile=0.13842745650867855,
    lower quartile=0.023742061858842905,
    upper whisker=0.30945713634171534,
    lower whisker=4.5468698248489334e-05,
    draw position=1.9000000000000001
  }] coordinates {};

  \addplot+[fill=color2, fill opacity=0.7, draw=black, boxplot prepared={
    median=0.03283971052196752,
    upper quartile=0.08578766814484752,
    lower quartile=0.012444955291758701,
    upper whisker=0.1957835539165391,
    lower whisker=3.078639299860984e-05,
    draw position=2.9000000000000004
  }] coordinates {};

  \addplot+[fill=color2, fill opacity=0.7, draw=black, boxplot prepared={
    median=0.017785930972475356,
    upper quartile=0.06144430745848417,
    lower quartile=0.006885333945973682,
    upper whisker=0.13427589139855284,
    lower whisker=1.145159536293144e-05,
    draw position=3.9000000000000004
  }] coordinates {};

  \addplot+[fill=color3, fill opacity=0.7, draw=black, boxplot prepared={
    median=0.01721296458809666,
    upper quartile=0.04935403537240882,
    lower quartile=0.00758641431136846,
    upper whisker=0.11162061690589897,
    lower whisker=3.0501368829282818e-05,
    draw position=0.09999999999999998
  }] coordinates {};

  \addplot+[fill=color3, fill opacity=0.7, draw=black, boxplot prepared={
    median=0.01473930509884323,
    upper quartile=0.040339032777543804,
    lower quartile=0.005738606205510777,
    upper whisker=0.09087011157942682,
    lower whisker=5.038077636749941e-05,
    draw position=1.1
  }] coordinates {};

  \addplot+[fill=color3, fill opacity=0.7, draw=black, boxplot prepared={
    median=0.00915855410743453,
    upper quartile=0.032550936468848504,
    lower quartile=0.003240069693410191,
    upper whisker=0.0762949833527468,
    lower whisker=1.7903522819169347e-05,
    draw position=2.1
  }] coordinates {};

  \addplot+[fill=color3, fill opacity=0.7, draw=black, boxplot prepared={
    median=0.005713855431458259,
    upper quartile=0.027310129629097665,
    lower quartile=0.002035406611926558,
    upper whisker=0.06491763465095646,
    lower whisker=2.2954570145878993e-06,
    draw position=3.1
  }] coordinates {};

  \addplot+[fill=color3, fill opacity=0.7, draw=black, boxplot prepared={
    median=0.004660244520702872,
    upper quartile=0.0460651675779617,
    lower quartile=0.0011123094904516547,
    upper whisker=0.11127635153426772,
    lower whisker=2.1102398377079316e-06,
    draw position=4.1
  }] coordinates {};

  \addplot+[fill=color4, fill opacity=0.7, draw=black, boxplot prepared={
    median=0.0030666260328681824,
    upper quartile=0.031865189059014254,
    lower quartile=0.0010524762098682334,
    upper whisker=0.07628249775525102,
    lower whisker=1.6372448700681405e-06,
    draw position=0.30000000000000004
  }] coordinates {};

  \addplot+[fill=color4, fill opacity=0.7, draw=black, boxplot prepared={
    median=0.002191171023441267,
    upper quartile=0.01931978977322408,
    lower quartile=0.0007028718770103103,
    upper whisker=0.04723479170123718,
    lower whisker=4.707305480551893e-06,
    draw position=1.3
  }] coordinates {};

  \addplot+[fill=color4, fill opacity=0.7, draw=black, boxplot prepared={
    median=0.0017631490704106446,
    upper quartile=0.019650391226794822,
    lower quartile=0.0004588994017221684,
    upper whisker=0.04843597293464621,
    lower whisker=1.8304267214006753e-06,
    draw position=2.3000000000000003
  }] coordinates {};

  \addplot+[fill=color4, fill opacity=0.7, draw=black, boxplot prepared={
    median=0.0025516510615886554,
    upper quartile=0.0317607608826013,
    lower quartile=0.0003289618583791817,
    upper whisker=0.07516190257686346,
    lower whisker=7.12973096386027e-07,
    draw position=3.3000000000000003
  }] coordinates {};

  \addplot+[fill=color4, fill opacity=0.7, draw=black, boxplot prepared={
    median=0.003096114987996359,
    upper quartile=0.08402180589970687,
    lower quartile=0.0002567081788224236,
    upper whisker=0.20449057333898252,
    lower whisker=3.9758984524010723e-07,
    draw position=4.3
  }] coordinates {};

\end{axis}
\end{tikzpicture}}
    }
    \caption{Velocity estimation results changing $T_{\max}$.}
    \label{fig:results_multiband_alg}
\vspace{-4mm}
\end{figure}
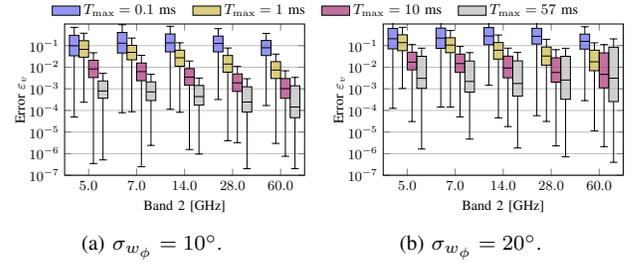

\textbf{Single component target.} \fig{fig:results_multiband_alg} shows estimation results under different parameter settings.
Comparing \fig{fig:results_multiband_alg_a} ($\sigma_{w_\phi} = 10^\circ$) with \fig{fig:results_multiband_alg_b} ($\sigma_{w_\phi} = 20^{\circ}$) shows a clear degradation with increasing noise. 
Notably, the estimation performance improves significantly as the frequency separation between $f_{1}$ and $f_{2}$ grows. 
This trend is attributed to the higher frequency diversity, which yields a more distinctive phase evolution across measurements. Moreover, the algorithm achieves superior results when \mbox{$T_{\max} = 57$~ms} is utilized, consistent with the Doppler \ac{crb}~\cite{dogandzic2001cramer},
which is inversely proportional to the observation time.
 

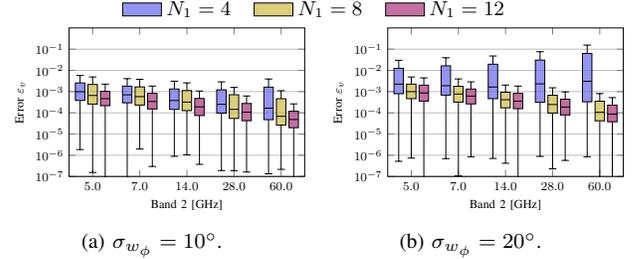
\begin{figure}[t!]
    \centering
    \subfloat{
        \resizebox{0.3\textwidth}{!}{\begin{tikzpicture}
\definecolor{color1}{RGB}{102,104,238}
\definecolor{color2}{RGB}{204,187,68}
\definecolor{color3}{RGB}{170,51,119}
\definecolor{color4}{RGB}{187,187,187}

\begin{axis}[
    width=0cm,
    height=0cm,
    axis line style={draw=none},
    tick style={draw=none},
    scale only axis,
    xmin=0, xmax=1,
    ymin=0, ymax=1,
    xtick=\empty,
    ytick=\empty,
    axis background/.style={fill=white},
    legend style={
        legend cell align=left,
        fill opacity=1,
        align=center,
        draw=white,
        font=\small,
        at={(0.5,0.5)},
        anchor=center,
        /tikz/every even column/.append style={column sep=0.8em} 
    },
    legend columns=4, 
]
\addlegendimage{area legend, fill=color1, fill opacity = 0.7, draw=black}
\addlegendentry{$N_1  = 4$}

\addlegendimage{area legend, fill=color2, draw=black, fill opacity = 0.7, }
\addlegendentry{$N_1  = 8$}

\addlegendimage{area legend, fill=color3, draw=black, fill opacity = 0.7}
\addlegendentry{$N_1  = 12$}
\end{axis}

\end{tikzpicture}}
    } \\
    \setcounter{subfigure}{0}
    \subfloat[$\sigma_{w_\phi} = 10^\circ$.  \label{fig:results_multiband_diff_pkts_10}]{
        \resizebox{0.22\textwidth}{!}{
\begin{tikzpicture}
\definecolor{color1}{RGB}{102,104,238}
\definecolor{color2}{RGB}{204,187,68}
\definecolor{color3}{RGB}{170,51,119}
\definecolor{color4}{RGB}{187,187,187}
\begin{axis}[
  boxplot/draw direction=y,
  ylabel={Error $\varepsilon_v$},
  xlabel={Band 2 [GHz]},
  ymin=1e-7, ymax=1,
  xmin=-0.5, xmax=4.5,
  ymode = log,
  yticklabel style={/pgf/number format/fixed},
  ymajorgrids,
  ytick={1e-1,1e-2,1e-3,1e-4,1e-5,1e-6,1e-7},
  xtick={0,1,2,3,4},   
  xticklabels={$5.0$,$7.0$,$14.0$,$28.0$,$60.0$},
  xlabel style={font=\footnotesize}, ylabel style={font=\footnotesize}, ticklabel style={font=\footnotesize},
  /pgfplots/boxplot/box extend=0.2,
  height=5cm,
  width=7cm,
  unbounded coords=discard,
  boxplot/every box/.style={solid, draw=black},
  boxplot/every whisker/.style={solid, black},
  boxplot/every cap/.style={solid, black},
  boxplot/every median/.style={solid, black},
]
  \addplot+[fill=color1, fill opacity=0.7, draw=black, boxplot prepared={
    median=0.0009819026227952075,
    upper quartile=0.0025581093948197583,
    lower quartile=0.000381414740142326,
    upper whisker=0.005777885395199722,
    lower whisker=1.8529345656090688e-06,
    draw position=-0.2666666666666667
  }] coordinates {};

  \addplot+[fill=color1, fill opacity=0.7, draw=black, boxplot prepared={
    median=0.0007038230083247589,
    upper quartile=0.001810502088922007,
    lower quartile=0.0002962313708487746,
    upper whisker=0.004081232631819968,
    lower whisker=8.37111251527717e-08,
    draw position=0.7333333333333333
  }] coordinates {};

  \addplot+[fill=color1, fill opacity=0.7, draw=black, boxplot prepared={
    median=0.00038104419095112056,
    upper quartile=0.001333600324788645,
    lower quartile=0.00014536012866703843,
    upper whisker=0.003100844035307913,
    lower whisker=8.894928355436591e-07,
    draw position=1.7333333333333334
  }] coordinates {};

  \addplot+[fill=color1, fill opacity=0.7, draw=black, boxplot prepared={
    median=0.0002502767793285374,
    upper quartile=0.0012162063011076476,
    lower quartile=9.692488547376097e-05,
    upper whisker=0.0028798010223701665,
    lower whisker=1.9030059745759361e-07,
    draw position=2.7333333333333334
  }] coordinates {};

  \addplot+[fill=color1, fill opacity=0.7, draw=black, boxplot prepared={
    median=0.00016615691256895024,
    upper quartile=0.0016235710889688677,
    lower quartile=4.7610128147144914e-05,
    upper whisker=0.003946353216644283,
    lower whisker=1.3336971406043792e-07,
    draw position=3.7333333333333334
  }] coordinates {};

  \addplot+[fill=color2, fill opacity=0.7, draw=black, boxplot prepared={
    median=0.000657636947199575,
    upper quartile=0.002170157611251242,
    lower quartile=0.0002589609477518839,
    upper whisker=0.004908771563569013,
    lower whisker=1.5464160171243465e-07,
    draw position=0.0
  }] coordinates {};

  \addplot+[fill=color2, fill opacity=0.7, draw=black, boxplot prepared={
    median=0.0005801161696165796,
    upper quartile=0.0016661484584290036,
    lower quartile=0.00023261920133373703,
    upper whisker=0.003809820097319316,
    lower whisker=1.9651008644432005e-06,
    draw position=1.0
  }] coordinates {};

  \addplot+[fill=color2, fill opacity=0.7, draw=black, boxplot prepared={
    median=0.000321560239228783,
    upper quartile=0.001121877860455863,
    lower quartile=0.00012725359084255557,
    upper whisker=0.0025794199442924232,
    lower whisker=1.0456455594022756e-06,
    draw position=2.0
  }] coordinates {};

  \addplot+[fill=color2, fill opacity=0.7, draw=black, boxplot prepared={
    median=0.00014582278399282253,
    upper quartile=0.0007124671355755748,
    lower quartile=5.472048083725165e-05,
    upper whisker=0.0015761830544086036,
    lower whisker=1.8791675500438224e-07,
    draw position=3.0
  }] coordinates {};

  \addplot+[fill=color2, fill opacity=0.7, draw=black, boxplot prepared={
    median=6.882626095681514e-05,
    upper quartile=0.0004621195151996876,
    lower quartile=2.6477744349723616e-05,
    upper whisker=0.0011027521600754332,
    lower whisker=2.1637148461124056e-07,
    draw position=4.0
  }] coordinates {};

  \addplot+[fill=color3, fill opacity=0.7, draw=black, boxplot prepared={
    median=0.0004648712998648612,
    upper quartile=0.0010319405770327103,
    lower quartile=0.00021675310596102433,
    upper whisker=0.0022536111531684783,
    lower whisker=1.115299817872452e-16,
    draw position=0.2666666666666666
  }] coordinates {};

  \addplot+[fill=color3, fill opacity=0.7, draw=black, boxplot prepared={
    median=0.0003416023944359265,
    upper quartile=0.0008291183839678921,
    lower quartile=0.0001516823877801677,
    upper whisker=0.0018242519153739735,
    lower whisker=2.963386062823726e-07,
    draw position=1.2666666666666666
  }] coordinates {};

  \addplot+[fill=color3, fill opacity=0.7, draw=black, boxplot prepared={
    median=0.000190856107797237,
    upper quartile=0.00047490283463918863,
    lower quartile=7.45505519804132e-05,
    upper whisker=0.001066227889252328,
    lower whisker=3.721799322479027e-07,
    draw position=2.2666666666666666
  }] coordinates {};

  \addplot+[fill=color3, fill opacity=0.7, draw=black, boxplot prepared={
    median=0.00010782919910476016,
    upper quartile=0.0002757852389204847,
    lower quartile=4.2636680714807326e-05,
    upper whisker=0.0006249448596863004,
    lower whisker=1.6199962270500466e-07,
    draw position=3.2666666666666666
  }] coordinates {};

  \addplot+[fill=color3, fill opacity=0.7, draw=black, boxplot prepared={
    median=4.850141943203468e-05,
    upper quartile=0.0001187629635066581,
    lower quartile=1.9945826846451004e-05,
    upper whisker=0.0002639111089185958,
    lower whisker=1.6199962270500466e-09,
    draw position=4.266666666666667
  }] coordinates {};

\end{axis}
\end{tikzpicture}}
    }
    \subfloat[$\sigma_{w_\phi} = 20^\circ$. \label{fig:results_multiband_diff_pkts_20}]{
        \resizebox{0.22\textwidth}{!}{
\begin{tikzpicture}
\definecolor{color1}{RGB}{102,104,238}
\definecolor{color2}{RGB}{204,187,68}
\definecolor{color3}{RGB}{170,51,119}
\definecolor{color4}{RGB}{187,187,187}
\begin{axis}[
  boxplot/draw direction=y,
  ylabel={Error $\varepsilon_v$},
  xlabel={Band 2 [GHz]},
  ymin=1e-7, ymax=1,
  xmin=-0.5, xmax=4.5,
  ymode = log,
  yticklabel style={/pgf/number format/fixed},
  ymajorgrids,
  ytick={1e-1,1e-2,1e-3,1e-4,1e-5,1e-6,1e-7},
  xtick={0,1,2,3,4},   
  xticklabels={$5.0$,$7.0$,$14.0$,$28.0$,$60.0$},
  xlabel style={font=\footnotesize}, ylabel style={font=\footnotesize}, ticklabel style={font=\footnotesize},
  /pgfplots/boxplot/box extend=0.2,
  height=5cm,
  width=7cm,
  boxplot/every box/.style={solid, draw=black},
  boxplot/every whisker/.style={solid, black},
  boxplot/every cap/.style={solid, black},
  boxplot/every median/.style={solid, black},
]
  \addplot+[fill=color1, fill opacity=0.7, draw=black, boxplot prepared={
    median=0.00225512639531712,
    upper quartile=0.012549298820053404,
    lower quartile=0.0007934325725999009,
    upper whisker=0.029887162534780683,
    lower whisker=5.181889110471313e-07,
    draw position=-0.2666666666666667
  }] coordinates {};

  \addplot+[fill=color1, fill opacity=0.7, draw=black, boxplot prepared={
    median=0.0018965711061297826,
    upper quartile=0.01636629009536368,
    lower quartile=0.0006840092277538783,
    upper whisker=0.039473217335975544,
    lower whisker=6.942215964224284e-07,
    draw position=0.7333333333333333
  }] coordinates {};

  \addplot+[fill=color1, fill opacity=0.7, draw=black, boxplot prepared={
    median=0.0016436693190974077,
    upper quartile=0.01954241754768633,
    lower quartile=0.0004646677143279369,
    upper whisker=0.04770395720694011,
    lower whisker=7.077786216000981e-07,
    draw position=1.7333333333333334
  }] coordinates {};

  \addplot+[fill=color1, fill opacity=0.7, draw=black, boxplot prepared={
    median=0.0023071127614350652,
    upper quartile=0.030655506926266136,
    lower quartile=0.00031368975009324134,
    upper whisker=0.07593773442398227,
    lower whisker=8.830634267497441e-07,
    draw position=2.7333333333333334
  }] coordinates {};

  \addplot+[fill=color1, fill opacity=0.7, draw=black, boxplot prepared={
    median=0.003052463596900221,
    upper quartile=0.06291605715834174,
    lower quartile=0.0003313501252098079,
    upper whisker=0.1558442767205784,
    lower whisker=8.510072426062407e-07,
    draw position=3.7333333333333334
  }] coordinates {};

  \addplot+[fill=color2, fill opacity=0.7, draw=black, boxplot prepared={
    median=0.0009943307561846451,
    upper quartile=0.0022517418465670238,
    lower quartile=0.0004643975347367811,
    upper whisker=0.0048914570703181975,
    lower whisker=7.367220961826729e-07,
    draw position=0.0
  }] coordinates {};

  \addplot+[fill=color2, fill opacity=0.7, draw=black, boxplot prepared={
    median=0.0007699683085983741,
    upper quartile=0.0017779480473923805,
    lower quartile=0.00031963744782728275,
    upper whisker=0.003963504656005717,
    lower whisker=1.067457966930269e-07,
    draw position=1.0
  }] coordinates {};

  \addplot+[fill=color2, fill opacity=0.7, draw=black, boxplot prepared={
    median=0.00041477168841314385,
    upper quartile=0.0009327333980610652,
    lower quartile=0.00017213053492963807,
    upper whisker=0.0020402831925435496,
    lower whisker=4.1900594967876627e-07,
    draw position=2.0
  }] coordinates {};

  \addplot+[fill=color2, fill opacity=0.7, draw=black, boxplot prepared={
    median=0.00024567396466525557,
    upper quartile=0.0006694228400601118,
    lower quartile=9.768840752912811e-05,
    upper whisker=0.001513673915278783,
    lower whisker=2.295163928322033e-07,
    draw position=3.0
  }] coordinates {};

  \addplot+[fill=color2, fill opacity=0.7, draw=black, boxplot prepared={
    median=0.00010556961211172398,
    upper quartile=0.0003529718632844284,
    lower quartile=4.215086643802409e-05,
    upper whisker=0.0008077566371739709,
    lower whisker=5.221144593390594e-08,
    draw position=4.0
  }] coordinates {};

  \addplot+[fill=color3, fill opacity=0.7, draw=black, boxplot prepared={
    median=0.0008566286432874052,
    upper quartile=0.002011279798380102,
    lower quartile=0.00035724935965613237,
    upper whisker=0.0044463635016127635,
    lower whisker=5.57208805153993e-08,
    draw position=0.2666666666666666
  }] coordinates {};

  \addplot+[fill=color3, fill opacity=0.7, draw=black, boxplot prepared={
    median=0.0006162727261895061,
    upper quartile=0.0013241236407333331,
    lower quartile=0.0002631070475761751,
    upper whisker=0.002891499121933403,
    lower whisker=8.506929943222535e-07,
    draw position=1.2666666666666666
  }] coordinates {};

  \addplot+[fill=color3, fill opacity=0.7, draw=black, boxplot prepared={
    median=0.00035336054013058126,
    upper quartile=0.0008275859872414751,
    lower quartile=0.00015694907253062628,
    upper whisker=0.0018234072686964691,
    lower whisker=6.078646606188834e-08,
    draw position=2.2666666666666666
  }] coordinates {};

  \addplot+[fill=color3, fill opacity=0.7, draw=black, boxplot prepared={
    median=0.00018678006716990988,
    upper quartile=0.000438240989226065,
    lower quartile=7.931009441127711e-05,
    upper whisker=0.0009760641139515655,
    lower whisker=5.68944475604639e-07,
    draw position=3.2666666666666666
  }] coordinates {};

  \addplot+[fill=color3, fill opacity=0.7, draw=black, boxplot prepared={
    median=8.625169816650624e-05,
    upper quartile=0.00023133188015813386,
    lower quartile=3.782722236091842e-05,
    upper whisker=0.0005194845347067607,
    lower whisker=9.753523022213223e-09,
    draw position=4.266666666666667
  }] coordinates {};

\end{axis}
\end{tikzpicture}}
    }
    \caption{Velocity estimation results changing the number of packets $N_1 = N_2$, with $T_{\max}=57$~ms.}
    \label{fig:results_multiband_diff_pkts}
\end{figure}

\fig{fig:results_multiband_diff_pkts} evaluates the impact of varying the number of packets $N_1 = N_2 \in \left \{ {4,8,12} \right \}$ using $T_{\max}=57$~ms. 
More packets quadratically increase the number of phase measurements, improving noise robustness and reducing the error below $10^{-3}$
for all carriers even at $\sigma_{w_\phi} = 20^\circ$.

\textbf{Multiple-component target.} In many \ac{isac} scenarios, targets consist of multiple reflecting components (e.g., drone blades, human limbs), each contributing a distinct Doppler shift.
We simulate a target with a dominant and a secondary moving components, using the same two-band configuration as in the previous sections, with $m=2$, $\tilde\beta_2 = 1 - \tilde\beta_1$, and $T_{\max}=57$~ms.
We estimate the velocity of component~$1$.

\fig{fig:multi_comp} shows the results for different values of $\tilde \beta_1$ to study the effect of varying the relative dominance of one component over the other. 
As expected, decreasing the value of $\tilde\beta_1$, thus increasing the relative weight of the second component, leads to performance degradation. 
Nevertheless, with moderately low carrier frequencies and $\sigma_{w_\phi} = 10^\circ$, the algorithm can accurately estimate the velocity for low $\tilde\beta_2$.



\begin{figure}[t!]
    \centering
    \subfloat{
        \resizebox{0.4\textwidth}{!}{\begin{tikzpicture}

\definecolor{color1}{RGB}{102,104,238}
\definecolor{color2}{RGB}{204,187,68}
\definecolor{color3}{RGB}{170,51,119}
\definecolor{color4}{RGB}{187,187,187}

\begin{axis}[
    width=0cm,
    height=0cm,
    axis line style={draw=none},
    tick style={draw=none},
    scale only axis,
    xmin=0, xmax=1,
    ymin=0, ymax=1,
    xtick=\empty,
    ytick=\empty,
    axis background/.style={fill=white},
    legend style={
        legend cell align=left,
        fill opacity=1,
        align=center,
        draw=white,
        font=\small,
        at={(0.5,0.5)},
        anchor=center,
        /tikz/every even column/.append style={column sep=0.8em} 
    },
    legend columns=4, 
]
\addlegendimage{area legend, fill=color1, fill opacity = 0.7, draw=black}
\addlegendentry{$\beta_1 = 0.9$}

\addlegendimage{area legend, fill=color2, draw=black, fill opacity = 0.7, }
\addlegendentry{$\beta_1 = 0.8$}

\addlegendimage{area legend, fill=color3, draw=black, fill opacity = 0.7}
\addlegendentry{$\beta_1 = 0.7$}

\addlegendimage{area legend, fill=color4, draw=black, fill opacity = 0.7}
\addlegendentry{$\beta_1 = 0.6$}

\end{axis}
\end{tikzpicture}}
    } \\
    \setcounter{subfigure}{0}
    \subfloat[$\sigma_{w_\phi} = 10^\circ$. \label{fig:multi_comp_10}]{
        \resizebox{0.22\textwidth}{!}{
\begin{tikzpicture}
\definecolor{color1}{RGB}{102,104,238}
\definecolor{color2}{RGB}{204,187,68}
\definecolor{color3}{RGB}{170,51,119}
\definecolor{color4}{RGB}{187,187,187}
\begin{axis}[
  boxplot/draw direction=y,
  ylabel={Error $\varepsilon_{v}$},
  xlabel={Band 2 [GHz]},
  ylabel shift=-5 pt,
  ymin=1e-7, ymax=1,
ytick={0.1, 0.01, 0.001, 0.0001, 0.00001, 0.000001, 0.0000001},
  ymode = log,
  xmin=-0.5, xmax=4.5,
  yticklabel style={/pgf/number format/fixed},
  ymajorgrids,
  xtick={0,1,2,3,4},
  xticklabels={$5.0$,$7.0$,$14.0$,$28.0$,$60.0$},
  xlabel style={font=\small}, ylabel style={font=\small}, ticklabel style={font=\small},
  /pgfplots/boxplot/box extend=0.2,
  height=5cm,
  width=7cm,
  boxplot/every box/.style={solid, draw=black},
  boxplot/every whisker/.style={solid, black},
  boxplot/every cap/.style={solid, black},
  boxplot/every median/.style={solid, black},
]
  \addplot+[fill=color1, fill opacity=0.7, draw=black, boxplot prepared={
    median=0.0010764190101104389,
    upper quartile=0.0026247833954493134,
    lower quartile=0.0004769921778474335,
    upper whisker=0.005812130992141702,
    lower whisker=2.9019539422538053e-06,
    draw position=-0.30000000000000004
  }] coordinates {};

  \addplot+[fill=color1, fill opacity=0.7, draw=black, boxplot prepared={
    median=0.0007771767491130978,
    upper quartile=0.0023256245553495694,
    lower quartile=0.00032265291267624685,
    upper whisker=0.0052962430236289096,
    lower whisker=6.451654240215373e-07,
    draw position=0.7
  }] coordinates {};

  \addplot+[fill=color1, fill opacity=0.7, draw=black, boxplot prepared={
    median=0.0004282898280450389,
    upper quartile=0.0015283496013141578,
    lower quartile=0.00015676163260160822,
    upper whisker=0.0035634206186835196,
    lower whisker=1.5950128059543322e-07,
    draw position=1.7000000000000002
  }] coordinates {};

  \addplot+[fill=color1, fill opacity=0.7, draw=black, boxplot prepared={
    median=0.0002748450686687598,
    upper quartile=0.0013450344108497446,
    lower quartile=9.600551972814174e-05,
    upper whisker=0.0032086655532827884,
    lower whisker=8.496557223287127e-09,
    draw position=2.7
  }] coordinates {};

  \addplot+[fill=color1, fill opacity=0.7, draw=black, boxplot prepared={
    median=0.0002081276005617006,
    upper quartile=0.0016578022749074012,
    lower quartile=5.682240681022135e-05,
    upper whisker=0.004005300855798656,
    lower whisker=4.403689590505381e-07,
    draw position=3.7
  }] coordinates {};

  \addplot+[fill=color2, fill opacity=0.7, draw=black, boxplot prepared={
    median=0.0016793459662924792,
    upper quartile=0.004902295191869888,
    lower quartile=0.0006567416249810599,
    upper whisker=0.011111055799031787,
    lower whisker=4.849382936422173e-06,
    draw position=-0.09999999999999998
  }] coordinates {};

  \addplot+[fill=color2, fill opacity=0.7, draw=black, boxplot prepared={
    median=0.0012656786503312187,
    upper quartile=0.004668756556857545,
    lower quartile=0.00046006257435317723,
    upper whisker=0.01002619954709074,
    lower whisker=1.1711505699003885e-06,
    draw position=0.9
  }] coordinates {};

  \addplot+[fill=color2, fill opacity=0.7, draw=black, boxplot prepared={
    median=0.0007298154236811637,
    upper quartile=0.004118920386813562,
    lower quartile=0.0002478770658422633,
    upper whisker=0.009900550087483147,
    lower whisker=1.8170409400887568e-07,
    draw position=1.9000000000000001
  }] coordinates {};

  \addplot+[fill=color2, fill opacity=0.7, draw=black, boxplot prepared={
    median=0.0005653450567037128,
    upper quartile=0.004751572130209799,
    lower quartile=0.00015704481288450044,
    upper whisker=0.011485493951209859,
    lower whisker=8.01365845222684e-07,
    draw position=2.9000000000000004
  }] coordinates {};

  \addplot+[fill=color2, fill opacity=0.7, draw=black, boxplot prepared={
    median=0.00047051744082400956,
    upper quartile=0.004191456865014794,
    lower quartile=9.825070973772184e-05,
    upper whisker=0.010255903576838263,
    lower whisker=3.6338079490768785e-08,
    draw position=3.9000000000000004
  }] coordinates {};

  \addplot+[fill=color3, fill opacity=0.7, draw=black, boxplot prepared={
    median=0.0030344203406881933,
    upper quartile=0.025409779671351624,
    lower quartile=0.0010733271555412427,
    upper whisker=0.06100150802918344,
    lower whisker=2.763831328614367e-07,
    draw position=0.09999999999999998
  }] coordinates {};

  \addplot+[fill=color3, fill opacity=0.7, draw=black, boxplot prepared={
    median=0.002527967537606354,
    upper quartile=0.02577209051752822,
    lower quartile=0.0009013512664003691,
    upper whisker=0.06214192308264124,
    lower whisker=4.230859926807752e-06,
    draw position=1.1
  }] coordinates {};

  \addplot+[fill=color3, fill opacity=0.7, draw=black, boxplot prepared={
    median=0.0022655769594136584,
    upper quartile=0.03337782904872425,
    lower quartile=0.0005368009042923583,
    upper whisker=0.08098106919084086,
    lower whisker=2.3334417154211895e-06,
    draw position=2.1
  }] coordinates {};

  \addplot+[fill=color3, fill opacity=0.7, draw=black, boxplot prepared={
    median=0.0032202705791934726,
    upper quartile=0.05359806102737753,
    lower quartile=0.0004238856222923992,
    upper whisker=0.1327736147311057,
    lower whisker=1.0104545146668877e-06,
    draw position=3.1
  }] coordinates {};

  \addplot+[fill=color3, fill opacity=0.7, draw=black, boxplot prepared={
    median=0.004135510900062108,
    upper quartile=0.10492611114292616,
    lower quartile=0.0004328027218920205,
    upper whisker=0.25668448427144736,
    lower whisker=1.2592939338113505e-06,
    draw position=4.1
  }] coordinates {};

  \addplot+[fill=color4, fill opacity=0.7, draw=black, boxplot prepared={
    median=0.031567975820409294,
    upper quartile=0.3632314250926203,
    lower quartile=0.0028172058697816644,
    upper whisker=0.8922844607853928,
    lower whisker=2.0587480696670455e-05,
    draw position=0.30000000000000004
  }] coordinates {};

  \addplot+[fill=color4, fill opacity=0.7, draw=black, boxplot prepared={
    median=0.03426113585316311,
    upper quartile=0.3884728462874901,
    lower quartile=0.0027132950985095327,
    upper whisker=0.9608858795244106,
    lower whisker=5.678202652282842e-07,
    draw position=1.3
  }] coordinates {};

  \addplot+[fill=color4, fill opacity=0.7, draw=black, boxplot prepared={
    median=0.051679842608801416,
    upper quartile=0.55202713863175,
    lower quartile=0.00453593605507583,
    upper whisker=1.3730900600915608,
    lower whisker=7.802543869935212e-06,
    draw position=2.3000000000000003
  }] coordinates {};

  \addplot+[fill=color4, fill opacity=0.7, draw=black, boxplot prepared={
    median=0.09288987167116593,
    upper quartile=0.7592745716586397,
    lower quartile=0.004811123993064063,
    upper whisker=1.88629743526366,
    lower whisker=7.397619736858937e-07,
    draw position=3.3000000000000003
  }] coordinates {};

  \addplot+[fill=color4, fill opacity=0.7, draw=black, boxplot prepared={
    median=0.12685170519459465,
    upper quartile=0.7486237352693392,
    lower quartile=0.006025152543606622,
    upper whisker=1.8498922067125794,
    lower whisker=8.001604795357663e-06,
    draw position=4.3
  }] coordinates {};
\end{axis}
\end{tikzpicture}}
    }
    \subfloat[$\sigma_{w_\phi} = 20^\circ$. \label{fig:multi_comp_20}]{
        \resizebox{0.22\textwidth}{!}{
\begin{tikzpicture}
\definecolor{color1}{RGB}{102,104,238}
\definecolor{color2}{RGB}{204,187,68}
\definecolor{color3}{RGB}{170,51,119}
\definecolor{color4}{RGB}{187,187,187}
\begin{axis}[
  boxplot/draw direction=y,
  ylabel={Error $\varepsilon_{v}$},
  xlabel={Band 2 [GHz]},
  ylabel shift=-5 pt,
  ymin=1e-7, ymax=1,
ytick={0.1, 0.01, 0.001, 0.0001, 0.00001, 0.000001, 0.0000001},
  ymode = log,
  xmin=-0.5, xmax=4.5,
  yticklabel style={/pgf/number format/fixed},
  ymajorgrids,
  xtick={0,1,2,3,4},
  xticklabels={$5.0$,$7.0$,$14.0$,$28.0$,$60.0$},
  xlabel style={font=\small}, ylabel style={font=\small}, ticklabel style={font=\small},
  /pgfplots/boxplot/box extend=0.2,
  height=5cm,
  width=7cm,
  boxplot/every box/.style={solid, draw=black},
  boxplot/every whisker/.style={solid, black},
  boxplot/every cap/.style={solid, black},
  boxplot/every median/.style={solid, black},
]
  \addplot+[fill=color1, fill opacity=0.7, draw=black, boxplot prepared={
    median=0.002769892969411122,
    upper quartile=0.024401626357406083,
    lower quartile=0.0010474312123274474,
    upper whisker=0.05899815384174816,
    lower whisker=9.523908335617427e-07,
    draw position=-0.30000000000000004
  }] coordinates {};

  \addplot+[fill=color1, fill opacity=0.7, draw=black, boxplot prepared={
    median=0.0024040025450733924,
    upper quartile=0.024407324063837794,
    lower quartile=0.0007964050718012696,
    upper whisker=0.05956272936853976,
    lower whisker=1.0853859075938203e-06,
    draw position=0.7
  }] coordinates {};

  \addplot+[fill=color1, fill opacity=0.7, draw=black, boxplot prepared={
    median=0.0018011298866123153,
    upper quartile=0.0224952042970668,
    lower quartile=0.0004485411725076565,
    upper whisker=0.055234699938983355,
    lower whisker=6.945732730969202e-07,
    draw position=1.7000000000000002
  }] coordinates {};

  \addplot+[fill=color1, fill opacity=0.7, draw=black, boxplot prepared={
    median=0.002856117066167739,
    upper quartile=0.047263110785923934,
    lower quartile=0.00031823543166982753,
    upper whisker=0.11379702016669602,
    lower whisker=1.2761897393040593e-06,
    draw position=2.7
  }] coordinates {};

  \addplot+[fill=color1, fill opacity=0.7, draw=black, boxplot prepared={
    median=0.003284290790874122,
    upper quartile=0.07425304574709184,
    lower quartile=0.0003011135758574218,
    upper whisker=0.18504587060655292,
    lower whisker=1.1819360245276328e-06,
    draw position=3.7
  }] coordinates {};

  \addplot+[fill=color2, fill opacity=0.7, draw=black, boxplot prepared={
    median=0.0038100641763966197,
    upper quartile=0.0445441453430618,
    lower quartile=0.0011300525996243969,
    upper whisker=0.10718338592545926,
    lower whisker=1.5781075450943094e-05,
    draw position=-0.09999999999999998
  }] coordinates {};

  \addplot+[fill=color2, fill opacity=0.7, draw=black, boxplot prepared={
    median=0.0033572564365371136,
    upper quartile=0.04366721294362484,
    lower quartile=0.0009914966598456847,
    upper whisker=0.10754270080794352,
    lower whisker=3.608152796313352e-06,
    draw position=0.9
  }] coordinates {};

  \addplot+[fill=color2, fill opacity=0.7, draw=black, boxplot prepared={
    median=0.0030751391801256255,
    upper quartile=0.04120734897912499,
    lower quartile=0.0005433796118071124,
    upper whisker=0.1010330884338391,
    lower whisker=2.0199890432273917e-07,
    draw position=1.9000000000000001
  }] coordinates {};

  \addplot+[fill=color2, fill opacity=0.7, draw=black, boxplot prepared={
    median=0.003983916455172172,
    upper quartile=0.08045518977609707,
    lower quartile=0.00041677274877537855,
    upper whisker=0.19995067908818562,
    lower whisker=3.6353433972119826e-06,
    draw position=2.9000000000000004
  }] coordinates {};

  \addplot+[fill=color2, fill opacity=0.7, draw=black, boxplot prepared={
    median=0.00621651626416538,
    upper quartile=0.15635881762441386,
    lower quartile=0.0006163516991870399,
    upper whisker=0.3735975535538941,
    lower whisker=1.669817496959156e-06,
    draw position=3.9000000000000004
  }] coordinates {};

  \addplot+[fill=color3, fill opacity=0.7, draw=black, boxplot prepared={
    median=0.010296715297807853,
    upper quartile=0.16354297560455855,
    lower quartile=0.0017591671477516417,
    upper whisker=0.39285554074371887,
    lower whisker=3.827867162674178e-06,
    draw position=0.09999999999999998
  }] coordinates {};

  \addplot+[fill=color3, fill opacity=0.7, draw=black, boxplot prepared={
    median=0.011243669914560155,
    upper quartile=0.13820538713057118,
    lower quartile=0.00152469594746881,
    upper whisker=0.3397672063142518,
    lower whisker=2.9404518976520154e-06,
    draw position=1.1
  }] coordinates {};

  \addplot+[fill=color3, fill opacity=0.7, draw=black, boxplot prepared={
    median=0.011896925721296943,
    upper quartile=0.2149585769482333,
    lower quartile=0.0011639879875349763,
    upper whisker=0.5263301592949753,
    lower whisker=2.557362437482229e-06,
    draw position=2.1
  }] coordinates {};

  \addplot+[fill=color3, fill opacity=0.7, draw=black, boxplot prepared={
    median=0.019685314138221,
    upper quartile=0.2814038516117192,
    lower quartile=0.0014104577266964278,
    upper whisker=0.6750657725392752,
    lower whisker=4.059201702998346e-06,
    draw position=3.1
  }] coordinates {};

  \addplot+[fill=color3, fill opacity=0.7, draw=black, boxplot prepared={
    median=0.034969891378023116,
    upper quartile=0.41265041983847517,
    lower quartile=0.002247303466686855,
    upper whisker=1.0158243228938322,
    lower whisker=6.7553729480993115e-06,
    draw position=4.1
  }] coordinates {};

  \addplot+[fill=color4, fill opacity=0.7, draw=black, boxplot prepared={
    median=0.08523140083918389,
    upper quartile=0.7524968041536908,
    lower quartile=0.005228555424144908,
    upper whisker=1.86928712311211,
    lower whisker=5.101539790550797e-06,
    draw position=0.30000000000000004
  }] coordinates {};

  \addplot+[fill=color4, fill opacity=0.7, draw=black, boxplot prepared={
    median=0.06938318379202163,
    upper quartile=0.6973525391225974,
    lower quartile=0.006019077096104238,
    upper whisker=1.719859256266643,
    lower whisker=2.1135505614949896e-05,
    draw position=1.3
  }] coordinates {};

  \addplot+[fill=color4, fill opacity=0.7, draw=black, boxplot prepared={
    median=0.13288080875485955,
    upper quartile=0.8747350469522944,
    lower quartile=0.007399609538460432,
    upper whisker=2.1719917923782326,
    lower whisker=1.6316956901725966e-05,
    draw position=2.3000000000000003
  }] coordinates {};

  \addplot+[fill=color4, fill opacity=0.7, draw=black, boxplot prepared={
    median=0.1490180157419575,
    upper quartile=0.8369410731781767,
    lower quartile=0.009675355935448843,
    upper whisker=2.0763216753715104,
    lower whisker=6.3562653767132816e-06,
    draw position=3.3000000000000003
  }] coordinates {};

  \addplot+[fill=color4, fill opacity=0.7, draw=black, boxplot prepared={
    median=0.20443936354351186,
    upper quartile=0.9995762072766637,
    lower quartile=0.018571777225732205,
    upper whisker=2.4615531817263125,
    lower whisker=1.4729365614743282e-06,
    draw position=4.3
  }] coordinates {};

\end{axis}
\end{tikzpicture}}
    }
    \caption{Velocity estimation results for a multiple-component target varying the coefficient $\beta_1$, with $T_{\max}=57$~ms.}
    \label{fig:multi_comp}
\end{figure}
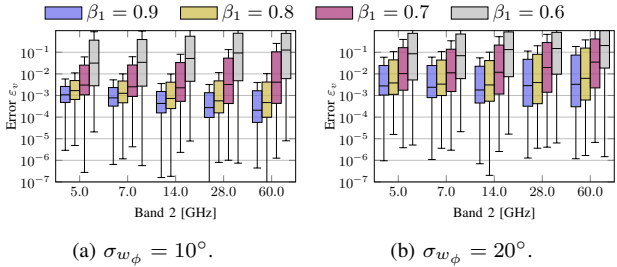

\begin{figure}[t!]
    \centering
    \subfloat{
        \resizebox{0.4\textwidth}{!}{\begin{tikzpicture}

\definecolor{color1}{RGB}{102,104,238}
\definecolor{color2}{RGB}{204,187,68}
\definecolor{color3}{RGB}{170,51,119}
\definecolor{color4}{RGB}{187,187,187}
\begin{axis}[
    width=0cm,
    height=0cm,
    axis line style={draw=none},
    tick style={draw=none},
    scale only axis,
    xmin=0, xmax=1,
    ymin=0, ymax=1,
    xtick=\empty,
    ytick=\empty,
    axis background/.style={fill=white},
    legend style={
        legend cell align=left,
        fill opacity=1,
        align=center,
        draw=white,
        font=\scriptsize,
        at={(0.5,0.5)},
        anchor=center,
        /tikz/every even column/.append style={column sep=1.5em} 
    },
    legend columns=3, 
]
\addlegendimage{area legend, fill=color1, draw=black, fill opacity = 0.7}
\addlegendentry{Multiband}

\addlegendimage{area legend, fill=color2, draw=black, fill opacity = 0.7}
\addlegendentry{IML}

\addlegendimage{area legend, fill=color3, draw=black, fill opacity = 0.7}
\addlegendentry{Single-band}

\end{axis}
\end{tikzpicture}}
    } \\
    \setcounter{subfigure}{0}
    \subfloat[$\sigma_{w_\phi} = 10^\circ$.\label{fig:results_comparisons_a}]{
        \resizebox{0.22\textwidth}{!}{
\begin{tikzpicture}
\definecolor{color1}{RGB}{102,104,238}
\definecolor{color2}{RGB}{204,187,68}
\definecolor{color3}{RGB}{170,51,119}
\definecolor{color4}{RGB}{187,187,187}
\begin{axis}[
  boxplot/draw direction=y,
  ylabel={Error $\varepsilon_v$},
  xlabel={Band 2 [GHz]},
  ylabel shift=-5 pt,
  ymin=1e-7, ymax=1,
ytick={0.1, 0.01, 0.001, 0.0001, 0.00001, 0.000001, 0.0000001},
  ymode = log,
  xmin=-0.5, xmax=4.5,
  yticklabel style={/pgf/number format/fixed},
  ymajorgrids,
  xtick={0,1,2,3,4},
  xticklabels={$5.0$,$7.0$,$14.0$,$28.0$,$60.0$},
  xlabel style={font=\small}, ylabel style={font=\small}, ticklabel style={font=\small},
  /pgfplots/boxplot/box extend=0.2,
  height=5cm,
  width=7cm,
  boxplot/every box/.style={solid, draw=black},
  boxplot/every whisker/.style={solid, black},
  boxplot/every cap/.style={solid, black},
  boxplot/every median/.style={solid, black},
]
  \addplot+[fill=color1, fill opacity=0.7, draw=black, boxplot prepared={
    median=0.0007942694126759071,
    upper quartile=0.002310815017750662,
    lower quartile=0.00036814208868943184,
    upper whisker=0.00522074573498139,
    lower whisker=5.1403664313985e-07,
    draw position=-0.375
  }] coordinates {};

  \addplot+[fill=color1, fill opacity=0.7, draw=black, boxplot prepared={
    median=0.0007225201706235215,
    upper quartile=0.002092558818822003,
    lower quartile=0.00028662163311118883,
    upper whisker=0.004720492067791785,
    lower whisker=2.3997484002769843e-06,
    draw position=0.625
  }] coordinates {};

  \addplot+[fill=color1, fill opacity=0.7, draw=black, boxplot prepared={
    median=0.00043611093067130307,
    upper quartile=0.0014316054065102069,
    lower quartile=0.00017637425006383306,
    upper whisker=0.003155745658605426,
    lower whisker=9.5521571244276e-07,
    draw position=1.625
  }] coordinates {};

  \addplot+[fill=color1, fill opacity=0.7, draw=black, boxplot prepared={
    median=0.00024406477969177447,
    upper quartile=0.0012414059614899305,
    lower quartile=8.396625581219836e-05,
    upper whisker=0.002942490776762759,
    lower whisker=2.0085657029566602e-07,
    draw position=2.625
  }] coordinates {};

  \addplot+[fill=color1, fill opacity=0.7, draw=black, boxplot prepared={
    median=0.00014401387271242364,
    upper quartile=0.0013826242918139036,
    lower quartile=4.611971216590009e-05,
    upper whisker=0.0033756822797758705,
    lower whisker=2.018114847447252e-07,
    draw position=3.625
  }] coordinates {};

  \addplot+[fill=color2, fill opacity=0.7, draw=black, boxplot prepared={
    median=0.0008064564265778327,
    upper quartile=0.0023292132105424503,
    lower quartile=0.00035523564751366115,
    upper whisker=0.005102446093471824,
    lower whisker=2.0057599190804676e-06,
    draw position=-0.125
  }] coordinates {};

  \addplot+[fill=color2, fill opacity=0.7, draw=black, boxplot prepared={
    median=0.0007533638532216246,
    upper quartile=0.002036194933487988,
    lower quartile=0.0002801392021543939,
    upper whisker=0.004637945785367045,
    lower whisker=2.5218855687150264e-06,
    draw position=0.875
  }] coordinates {};

  \addplot+[fill=color2, fill opacity=0.7, draw=black, boxplot prepared={
    median=0.001955123201731937,
    upper quartile=0.36588575490187114,
    lower quartile=0.0004960659723250618,
    upper whisker=0.9111050008533143,
    lower whisker=1.0270407307934376e-06,
    draw position=1.875
  }] coordinates {};

  \addplot+[fill=color2, fill opacity=0.7, draw=black, boxplot prepared={
    median=0.4436670157003714,
    upper quartile=0.934981401334873,
    lower quartile=0.0010730799320496501,
    upper whisker=1.8704186392445776,
    lower whisker=2.087026101199541e-06,
    draw position=2.875
  }] coordinates {};

  \addplot+[fill=color2, fill opacity=0.7, draw=black, boxplot prepared={
    median=0.8712407544462826,
    upper quartile=1.078746068475056,
    lower quartile=0.06714195051127358,
    upper whisker=1.8990771067036358,
    lower whisker=8.140685828773626e-06,
    draw position=3.875
  }] coordinates {};

  \addplot+[fill=color3, fill opacity=0.7, draw=black, boxplot prepared={
    median=0.00335515452364276,
    upper quartile=0.03757976069581621,
    lower quartile=0.0007051379032321756,
    upper whisker=0.09236435494576621,
    lower whisker=1.5861462471576507e-06,
    draw position=0.125
  }] coordinates {};

  \addplot+[fill=color3, fill opacity=0.7, draw=black, boxplot prepared={
    median=0.002341394818298224,
    upper quartile=0.025684554247465036,
    lower quartile=0.0004811273884132495,
    upper whisker=0.06287920602152834,
    lower whisker=2.6767671522162875e-06,
    draw position=1.125
  }] coordinates {};

  \addplot+[fill=color3, fill opacity=0.7, draw=black, boxplot prepared={
    median=0.002273919723978412,
    upper quartile=0.04323016086346352,
    lower quartile=0.0003304743448390049,
    upper whisker=0.107473176339319,
    lower whisker=8.241120864671574e-07,
    draw position=2.125
  }] coordinates {};

  \addplot+[fill=color3, fill opacity=0.7, draw=black, boxplot prepared={
    median=0.0030207809201674113,
    upper quartile=0.05112082632758159,
    lower quartile=0.0002015052669792927,
    upper whisker=0.12736069665070204,
    lower whisker=2.2666720065358982e-07,
    draw position=3.125
  }] coordinates {};

  \addplot+[fill=color3, fill opacity=0.7, draw=black, boxplot prepared={
    median=0.005813321345257938,
    upper quartile=0.1444942267893748,
    lower quartile=0.0001763688492317015,
    upper whisker=0.357352909042515,
    lower whisker=2.5215984052854724e-07,
    draw position=4.125
  }] coordinates {};

\end{axis}
\end{tikzpicture}}
    }
    \subfloat[$\sigma_{w_\phi} = 20^\circ$. \label{fig:results_comparisons_b}]{
        \resizebox{0.22\textwidth}{!}{
\begin{tikzpicture}
\definecolor{color1}{RGB}{102,104,238}
\definecolor{color2}{RGB}{204,187,68}
\definecolor{color3}{RGB}{170,51,119}
\definecolor{color4}{RGB}{187,187,187}
\begin{axis}[
  boxplot/draw direction=y,
  ylabel={Error $\varepsilon_v$},
  xlabel={Band 2 [GHz]},
  ylabel shift=-5 pt,
  ymin=1e-7, ymax=1,
ytick={0.1, 0.01, 0.001, 0.0001, 0.00001, 0.000001, 0.0000001},
ymode=log,
  xmin=-0.5, xmax=4.5,
  yticklabel style={/pgf/number format/fixed},
  ymajorgrids,
  xtick={0,1,2,3,4},
  xticklabels={$5.0$,$7.0$,$14.0$,$28.0$,$60.0$},
  xlabel style={font=\small}, ylabel style={font=\small}, ticklabel style={font=\small},
  /pgfplots/boxplot/box extend=0.2,
  height=5cm,
  width=7cm,
  boxplot/every box/.style={solid, draw=black},
  boxplot/every whisker/.style={solid, black},
  boxplot/every cap/.style={solid, black},
  boxplot/every median/.style={solid, black},
]
  \addplot+[fill=color1, fill opacity=0.7, draw=black, boxplot prepared={
    median=0.0030666260328681824,
    upper quartile=0.031865189059014254,
    lower quartile=0.0010524762098682334,
    upper whisker=0.07628249775525102,
    lower whisker=1.6372448700681405e-06,
    draw position=-0.375
  }] coordinates {};

  \addplot+[fill=color1, fill opacity=0.7, draw=black, boxplot prepared={
    median=0.002191171023441267,
    upper quartile=0.01931978977322408,
    lower quartile=0.0007028718770103103,
    upper whisker=0.04723479170123718,
    lower whisker=4.707305480551893e-06,
    draw position=0.625
  }] coordinates {};

  \addplot+[fill=color1, fill opacity=0.7, draw=black, boxplot prepared={
    median=0.0017631490704106446,
    upper quartile=0.019650391226794822,
    lower quartile=0.0004588994017221684,
    upper whisker=0.04843597293464621,
    lower whisker=1.8304267214006753e-06,
    draw position=1.625
  }] coordinates {};

  \addplot+[fill=color1, fill opacity=0.7, draw=black, boxplot prepared={
    median=0.0025516510615886554,
    upper quartile=0.0317607608826013,
    lower quartile=0.0003289618583791817,
    upper whisker=0.07516190257686346,
    lower whisker=7.12973096386027e-07,
    draw position=2.625
  }] coordinates {};

  \addplot+[fill=color1, fill opacity=0.7, draw=black, boxplot prepared={
    median=0.003096114987996359,
    upper quartile=0.08402180589970687,
    lower quartile=0.0002567081788224236,
    upper whisker=0.20449057333898252,
    lower whisker=3.9758984524010723e-07,
    draw position=3.625
  }] coordinates {};

  \addplot+[fill=color2, fill opacity=0.7, draw=black, boxplot prepared={
    median=0.0028659389691353017,
    upper quartile=0.02696100108018894,
    lower quartile=0.0010062925020620308,
    upper whisker=0.06413597707626549,
    lower whisker=3.2610281566922494e-07,
    draw position=-0.125
  }] coordinates {};

  \addplot+[fill=color2, fill opacity=0.7, draw=black, boxplot prepared={
    median=0.002033089680190329,
    upper quartile=0.016487600201536178,
    lower quartile=0.0007037955687448967,
    upper whisker=0.04002696352582897,
    lower whisker=2.8419454979984814e-08,
    draw position=0.875
  }] coordinates {};

  \addplot+[fill=color2, fill opacity=0.7, draw=black, boxplot prepared={
    median=0.01501333282589473,
    upper quartile=0.5582298159883086,
    lower quartile=0.0013440500112481223,
    upper whisker=1.392785021376017,
    lower whisker=4.64661967285223e-06,
    draw position=1.875
  }] coordinates {};

  \addplot+[fill=color2, fill opacity=0.7, draw=black, boxplot prepared={
    median=0.6358439324263182,
    upper quartile=1.0023642698410808,
    lower quartile=0.004840000827612259,
    upper whisker=2.384728284659061,
    lower whisker=3.8208179835413654e-07,
    draw position=2.875
  }] coordinates {};

  \addplot+[fill=color2, fill opacity=0.7, draw=black, boxplot prepared={
    median=0.8720993394066252,
    upper quartile=1.0509994185472789,
    lower quartile=0.24750531451083235,
    upper whisker=2.0285366637312876,
    lower whisker=2.5861827484476853e-05,
    draw position=3.875
  }] coordinates {};

  \addplot+[fill=color3, fill opacity=0.7, draw=black, boxplot prepared={
    median=0.05623607240419739,
    upper quartile=0.37140861715283335,
    lower quartile=0.005517076909456735,
    upper whisker=0.9187345710837206,
    lower whisker=2.585453952423433e-05,
    draw position=0.125
  }] coordinates {};

  \addplot+[fill=color3, fill opacity=0.7, draw=black, boxplot prepared={
    median=0.051596244848781045,
    upper quartile=0.2982883163113844,
    lower quartile=0.003825480716135884,
    upper whisker=0.7318271253236616,
    lower whisker=1.730771874584294e-05,
    draw position=1.125
  }] coordinates {};

  \addplot+[fill=color3, fill opacity=0.7, draw=black, boxplot prepared={
    median=0.060459531986605024,
    upper quartile=0.26258143926562183,
    lower quartile=0.005324230243812553,
    upper whisker=0.6468189621481213,
    lower whisker=1.3373084111112288e-05,
    draw position=2.125
  }] coordinates {};

  \addplot+[fill=color3, fill opacity=0.7, draw=black, boxplot prepared={
    median=0.061739080494801354,
    upper quartile=0.23015381377877234,
    lower quartile=0.003575688814337335,
    upper whisker=0.5624306814854414,
    lower whisker=1.8151123958360018e-06,
    draw position=3.125
  }] coordinates {};

  \addplot+[fill=color3, fill opacity=0.7, draw=black, boxplot prepared={
    median=0.07802100032483791,
    upper quartile=0.3049111476727883,
    lower quartile=0.005582650327110875,
    upper whisker=0.7504620398262679,
    lower whisker=1.530470972642246e-07,
    draw position=4.125
  }] coordinates {};

\end{axis}
\end{tikzpicture}}
    }
    \caption{Velocity estimation results comparisons, with $T_{\max}=57$~ms.}
    \label{fig:results_comparisons}
\end{figure}
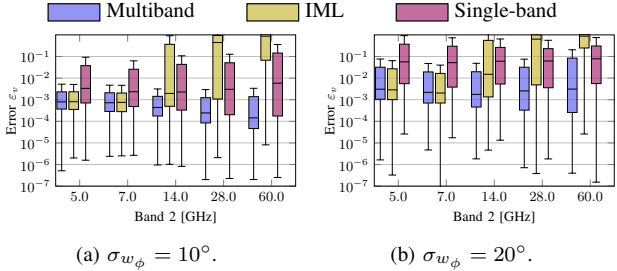 

\textbf{Comparison with existing methods.} We now evaluate our method against two benchmarks, on a single-component target.
\begin{itemize}
    \item \textit{Iterative maximum likelihood (IML).} We adapt the carrier phase multiband localization method in~\cite{bedin2025millimeter}, to solve our velocity estimation problem.
    This approach computes the product of the likelihood functions of the ambiguous phase measurements in the different bands.
    The likelihood in lower frequency bands exhibits wide peaks but has low ambiguity, while that at higher frequencies shows narrow peaks (high accuracy) but also high ambiguity.
    \item \textit{Single-band integer least-squares.} We also consider a baseline scenario with a single-band system working at $f_2$ (the highest carrier frequency), which is affected by ambiguity when the \ac{tdoa} is large.
\end{itemize}

The results of this comparison, are illustrated in \fig{fig:results_comparisons} across both noise levels.
The proposed multiband algorithm consistently achieves the lowest relative error. 
The single-band method suffers from high variance due to the scarce availability of unambiguous packet pairs, worsening at $60$~GHz.

Meanwhile, IML frequently locks onto incorrect integer ambiguities for $f_2\geq 14$~GHz because multiple peaks of the high-frequency likelihood fall within the main peak of the low-frequency one, limiting IML to cases with moderate frequency separation.
In contrast, our multiband approach leverages the full set of opportunistic packet arrivals at multiple carrier frequencies without constraints, even when the carriers are very different, thus providing accurate, robust, and unambiguous velocity estimation.

\section{Concluding remarks}\label{sec:conclusion}

In this paper, we have addressed the critical challenge of Doppler ambiguity in \ac{isac} systems where channel estimation is opportunistic and dictated by irregular communication patterns.
To overcome the limitations of traditional velocity estimation methods, we proposed an algorithm that leverages multiple frequency bands and all available inter-packet arrival times, solving a mixed-integer quadratic program to resolve integer phase ambiguities. 
This approach eliminates the need for dedicated sensing packets or high-density traffic injection, achieving a truly zero-overhead sensing solution. 
Our numerical evaluations demonstrate that the proposed method significantly outperforms multiband maximum likelihood-based and single-band benchmarks. 

\bibliography{references.bib}
\bibliographystyle{IEEEtran}
\end{document}